\documentclass{aa} 
\newcommand{\gtsim}{\protect\raisebox{-0.5ex}{$\:\stackrel{\textstyle >}
        {\sim}\:$}}
\newcommand{\ltsim}{\protect\raisebox{-0.5ex}{$\:\stackrel{\textstyle <}
        {\sim}\:$}}
\newcommand{\teff}{$T_{\rm eff}$}
\newcommand{\logg}{$\log g$}
\newcommand{\logeps}{$\log \epsilon$}
\newcommand{\csi}{$\xi$}
\newcommand{\feh}{[Fe/H]} 
\newcommand{\tih}{[Ti/H]} 
\newcommand{\xh}{[X/H]}
\newcommand{\vsini}{$v\sin i$}

\usepackage[flushleft]{threeparttable}

\usepackage{graphicx}
\usepackage{hyperref}
\usepackage{txfonts} 
\usepackage{natbib}

\usepackage{orcidlink}

\usepackage{multirow, makecell}
\definecolor{antiquewhite}{rgb}{0.98, 0.92, 0.84}
\definecolor{magnolia}{rgb}{0.97, 0.96, 1.0}
\bibpunct{(}{)}{;}{a}{}{,} 
\makeatother

\begin{document}

\title{The GAPS Programme at TNG}
\subtitle{LXI. Atmospheric parameters and elemental abundances of TESS young exoplanet host stars\thanks{Based on observations made with the Italian Telescopio Nazionale {\it Galileo} (TNG), operated on the island of La Palma by the INAF - Fundaci\'on Galileo Galilei at the Roque de los Muchachos Observatory of the Instituto de Astrof\'isica de Canarias (IAC) in the 
framework of the large programme Global Architecture of Planetary Systems (GAPS; PI: G. Micela).}}
   
\author{
S. Filomeno\inst{\ref{oaroma}, \ref{unitor}, \ref{sapienza}} \orcidlink{0009-0000-5623-5237} \and
K. Biazzo\inst{\ref{oaroma}} \orcidlink{0000-0002-1023-6821} \and
M. Baratella\inst{\ref{esochile}}\orcidlink{0000-0002-1027-5003} \and
S. Benatti\inst{\ref{oapalermo}} \and
V. D'Orazi\inst{\ref{unitor}, \ref{oapadova}} \and
S. Desidera\inst{\ref{oapadova}} \and
L.~Mancini\inst{\ref{unitor}, \ref{oatorino}, \ref{mpia}} \and
S.~Messina\inst{\ref{oacatania}}\and
D.~Polychroni\inst{\ref{oatrieste}} \and
D.~Turrini\inst{\ref{oatorino}, \ref{iaps}} \and
L. Cabona\inst{\ref{oapadova}} \and
I. Carleo\inst{\ref{iac}, \ref{ull}, \ref{oatorino}} \and
M. Damasso\inst{\ref{oatorino}} \and
L. Malavolta\inst{\ref{oapadova}, \ref{unipadova}} \and
G. Mantovan\inst{\ref{unipadova}, \ref{oapadova}} \and
D.~Nardiello\inst{\ref{unipadova}, \ref{oapadova}} \and
G. Scandariato\inst{\ref{oacatania}}\and
A. Sozzetti\inst{\ref{oatorino}} \and
T.~Zingales\inst{\ref{oapadova}, \ref{unipadova}} 
\and
G.~Andreuzzi\inst{\ref{tng}, \ref{oaroma}}\and S.~Antoniucci\inst{\ref{oaroma}}\and
A.~Bignamini\inst{\ref{oatrieste}} \and
A. S. Bonomo\inst{\ref{oatorino}} \and
R.~Claudi\inst{\ref{oapadova}, \ref{uniroma3}} \and
R.~Cosentino\inst{\ref{tng}} \and
A. F. M. Fiorenzano\inst{\ref{tng}}\orcidlink{0000-0002-4272-4272} \and
S.~Fonte\inst{\ref{iaps}} \and
A.~Harutyunyan\inst{\ref{tng}} \and
C.~Knapic\inst{\ref{oatrieste}} 
}
\offprints{S. Filomeno}
\mail{simone.filomeno@inaf.it}
\institute{
INAF-Osservatorio Astronomico di Roma, Via Frascati 33, I-00040 Monte Porzio Catone (RM), Italy \label{oaroma} \\
\email{simone.filomeno@inaf.it}
\and Dipartimento di Fisica, Universit\`a di Roma Tor Vergata, Via della Ricerca Scientifica 1, I-00133 Roma, Italy \label{unitor}
\and Dipartimento di Fisica, Sapienza Università di Roma, Piazzale Aldo Moro 5, I-00185 Roma, Italy \label{sapienza}
\and ESO - European Southern Observatory, Alonso de Córdova 3107, Casilla 19, Santiago 19001, Chile \label{esochile}
\and INAF - Osservatorio Astronomico di Palermo, Piazza del Parlamento 1, I-90134 Palermo, Italy \label{oapalermo}
\and INAF - Osservatorio Astronomico di Padova, Vicolo dell'Osservatorio 5, I-35122 Padova, Italy \label{oapadova}
\and INAF - Osservatorio Astrofisico di Torino, Via Osservatorio 20, I-10025 Pino Torinese, Italy, \label{oatorino}
\and Max Planck Institute for Astronomy, K\"{o}nigstuhl 17, D-69117, Heidelberg, Germany \label{mpia}
\and INAF - Osservatorio Astrofisico di Catania, Via S. Sofia 78, I-95123 Catania, Italy \label{oacatania}
\and INAF - Osservatorio Astronomico di Trieste, Via Tiepolo 11, I-34143 Trieste, Italy \label{oatrieste}
\and INAF - Istituto di Astrofisica e Planetologia Spaziali, Via del Fosso del Cavaliere 100, I-00133 Roma, Italy \label{iaps}
\and Instituto de Astrofísica de Canarias (IAC), Calle Vía Láctea s/n, 38200 La Laguna, Tenerife, Spain \label{iac}
\and Departamento de Astrofísica, Universidad de La Laguna (ULL), 38206 La Laguna, Tenerife, Spain \label{ull}
\and Dipartimento di Fisica e Astronomia, Universit$\grave{\rm a}$ di Padova, Vicolo dell'Osservatorio 2, I-35122 Padova, Italy \label{unipadova}
\and Fundaci{\'o}n Galileo Galilei - INAF, Rambla Jos{\'e} Ana Fernandez P{\'e}rez 7, E-38712 Bre$\tilde{\rm n}$a Baja (TF), Spain \label{tng} 
\and Dipartimento di Matematica e Fisica, Università Roma Tre, Via della Vasca Navale 84, 00146 Roma, Italy \label{uniroma3}
}

\date{Received .../ accepted ...}

\abstract 
{The study of exoplanets at different evolutionary stages can shed light on their formation, migration, and evolution. The determination of exoplanet properties depends on the properties of their host stars. It is therefore important to characterise the host stars for accurate knowledge on their planets.
}
{Our final goal is to derive, in a homogeneous and accurate way, the stellar atmospheric parameters and elemental abundances of ten young TESS (Transiting Exoplanet Survey Satellite) transiting planet-hosting GK stars followed up with the HARPS-N (High Accuracy Radial velocity Planet Searcher for the Northern hemisphere) at TNG (Telescopio Nazionale
\textit{Galileo}) spectrograph within the Global Architecture of Planetary Systems (GAPS) programme.
}
{We derived stellar kinematic properties, atmospheric parameters, and abundances of 18 elements. Depending on stellar parameters and chemical elements, we used methods based on line equivalent widths and spectral synthesis. Lithium line measurements were used as approximate age estimations. We exploited chemical abundances and their ratios to derive information on planetary composition.
}
{Elemental abundances and kinematic properties are consistent with the nearby Galactic thin disk. All targets show C/O<0.8 and 1.0<Mg/Si<1.5, compatible with an interpretation of silicate mantles made of a mixture of pyroxene and olivine assemblages for any rocky planets around them. The Fe/Mg ratios, with values of $\sim$0.7-1.0, show a propensity for the planets to have big (iron) cores. All stars hosting very low-mass planets 
show Mg/Si values consistent with the Earth values, thus demonstrating their similar mantle composition. Hot Jupiter host stars show a lower content of O/Si, which could be related to the lower presence of water content. We confirm a trend found in the literature between stellar [O/Fe] and total planetary mass, implying an important role of the O in shaping the mass fraction of heavy elements in stars and their disks.
}
{Exploring the composition of planets through the use of elemental abundances of their hosting stars holds promise for future investigations, particularly with large samples. Meanwhile, the detailed host star abundances provided can be employed for further studies on the composition of the planets within the current sample, when their atmospheres will be exploited.
}

           
\titlerunning{Atmospheric parameters and abundances of TESS planet host stars}
\authorrunning{S. Filomeno et al.}
\maketitle

\section{Introduction}
\label{sec:intro}

Precise stellar parameters and elemental abundances of planet-hosting stars are crucial for correctly interpreting exoplanet observations. Ground-based high-resolution spectroscopy provides the main avenue for determining the stellar parameters of exoplanet hosts. In recent years, much effort by many researchers in the community has pushed this work forward. Most of them are focussed on old FGK (\citealt{Sousaetal2021}, \citealt{Magrinietal2022}, \citealt{Biazzoetal2022}, \citealt{MacDougalletal2023}, \citealt{Strassmeier_2023}, \citealt{yun_2024}) and M (\citealt{Maldonadoetal2020}, \citealt{Marfiletal2021}) dwarf planet-hosting stars. In this context, the characterisation of transiting exoplanet host stars is of particular interest because they represent unique laboratories to test theories of exoplanet formation and evolution with a relatively high precision and possibly trace any potential migration they underwent (\citealt{Biazzoetal2022}, \citealt{KoleckiWang2022}). 

Planet formation models and theories can be constrained through the properties of the stellar atmospheres, as the composition of the host star can be used as a proxy for the composition of the initial protoplanetary disk from which the planets were born. Therefore, by comparing planetary and stellar atmospheric chemical composition, we can gain insight into where and how planets formed and possibly from where they migrated. This has, in turn, led modelling studies to focus on predicting the enrichment of volatiles, such as carbon and oxygen, in giant planet atmospheres (\citealt{Obergetal2011},\citealt{Madhusudhanetal2014},\citealt{Fonteetal2023}). Although this framework provides useful constraints on the potential formation location of gas giant exoplanets, carbon and oxygen measurements alone are not enough to determine from where a given gas giant planet originates. The degeneracy in the information provided by, for example, the C/O ratio can be broken by characterising the abundances of both volatile elements (such as sulphur, in addition to carbon and oxygen) and refractory elements (such as magnesium and silicon), which condense at higher temperatures. This can provide stronger constraints on the formation region and extent of planet migration. Therefore, the ratio of refractory content relative to volatile content in the atmosphere can give us direct insights into the nature of the solids that were incorporated into the planetary envelope (\citealt{Turrini2021,Turrini2022}; \citealt{Pacettietal2022}). This is particularly interesting for young stars because, compared with older counterparts that possibly experienced migration throughout the Galactic disk, their chemical content should have preserved much of the original chemical composition of the cloud from which they formed (\citealt{Spitonietal2023}, \citealt{Magrinietal2023}). Therefore, the chemistry of young exoplanet-hosting stars, and thus of their planets, is strongly connected to the time and place of its birth, unless the gas from which they formed has undergone peculiar enrichment.

In this paper, we conduct a homogeneous and accurate stellar parameter and elemental abundance analysis of several atomic species, including volatiles and refractories (such as carbon, oxygen, sulphur, magnesium, silicon, titanium, and iron), which are processed in planetary formation and migration. The analysis was performed as in previous works (\citealt{Biazzoetal2022}) in the context of the Italian Global Architecture of Planetary Systems (GAPS) programme \citep{covino_gaps_2013} and within the sub-programme devoted to relatively young transiting planet host stars (\citealt{baratellaetal2020b}, \citealt{carleo2020}). Inside the whole sample of targets in this sub-programme, we studied a subset of ten relatively young dwarf stars for which no similar extensive analysis has been previously done involving all elements we measured. Our final aim is to provide accurate stellar parameters and abundances of multiple elements using high-quality spectroscopic data, to possibly link stellar elemental abundances to the composition of their planets, and to investigate possible relationships between stellar abundances and the properties of their transiting companions. This will represent a step forward in providing necessary information for future studies on the atmospheric abundance of relatively young planets with facilities such as \textit{JWST} (James Webb Space Telescope) \citep{gardner_jwst_2023} and the next-coming \textit{Ariel} (Atmospheric Remote-Sensing Infrared Exoplanet Large-survey) telescope \citep{Tinettietal2021}.

The paper is organised as follows. In Sects.\,\ref{sec:obs} and \ref{sec:data_analysis} we describe the spectroscopic dataset and the measurements of stellar parameters and elemental abundances. In Sect.\,\ref{sec:result_discussion} we present and discuss our results in the context of chemical evolution, kinematics, and planetary abundances and properties. Finally, in Sect.\,\ref{sec:conclusions} we draw our conclusions.

\section{Stellar sample, observations, and data reduction}
\label{sec:obs}

The GAPS-Young Objects (YOs) sub-programme has the goal to study young and intermediate-aged planetary systems (with typical ages of $\leq$1\,Gyr) and to characterise their exoplanets down to sub-Neptune masses with transit and radial velocity analysis techniques, after accurate determinations of the host star properties. Within the GAPS-YOs sample, we selected TESS (Transiting Exoplanet Survey Satellite) Objects of Interest (TOIs; \citealt{ricker_2014}) hosting at least one known transiting planet and showing indications of young ages from our systematic check of their stellar properties (see \citealt{carleo2020}).
Then, within this sample, we selected targets with effective temperatures (\teff) higher than 5000\,K and rotational velocities ($v\sin i$) lower than 10\,km/s known from the NASA Exoplanet Archive. This was done to avoid problems caused by the presence of molecular lines or by line blending, which need a different approach than that adopted in this work (\citealt{Biazzoetal2022}, and references therein). We also considered two targets not belonging to the GAPS-YOs sub-programme but with similar characteristics and observed within the GAPS collaboration (see Table \ref{tab:target_references}). The final sample consists of ten targets with $V$ magnitude between $\sim$6.9 and $\sim$12.0 mag. All these targets have one to six confirmed planets with masses ranging from $\sim 2.4 M_{\rm E}$ to $\sim 10 M_{\rm J}$ (rocky to giant planets) as shown in more detail in the discovery papers mentioned in Tables\,\ref{tab:target_references}. 

Observations were performed between the beginning of 2020 and the end of 2023 with the High Accuracy Radial velocity Planet Searcher for the Northern hemisphere (HARPS-N; resolving power $R \sim 115000$, spectral range $\lambda \sim$ 3830-6900\,\AA; \citealt{cosentinoetal2012}) placed at the 3.6\,m Telescopio Nazionale {\it Galileo} (TNG) in La Palma, Spain. Spectra reduction was performed by using the standard HARPS-N Data Reduction Software \citep{pepe_harps_2022} and all the spectra collected for each target were subsequently combined to obtain a co-added spectrum with signal-to-noise ratio $S/N$ > 100 at $\sim$6000\,\AA.

Since we are interested in the oxygen abundance and one of its diagnostics is the near-IR triplet at $\sim$7770\,\AA, that is outside the HARPS-N spectral range, we searched for spectra acquired with the Tillinghast Reflector Echelle Spectrograph (TRES; $R \sim 44000$, $\lambda \sim$ 3850-9100\,\AA; \citealt{Szentgyorgyi_furesz_2007}) spectrograph at the Tillinghast telescope (Mt. Hopkins, Arizona) or with the CHIRON ($R \sim 80000$, $\lambda \sim$ 4150-8800\,\AA; \citealt{chiron_tokovinin2013}) spectrograph at the Cerro Tololo Inter-American Observatory (Chile) and available in the Exoplanet Follow-up Observing Programme (ExoFOP) website. In the end, we found TRES spectra for eight stars (TOI-1430, TOI-1726, TOI-2076, TOI-4515, TOI-5082, TOI-5398, TOI-5543, TOI-1136) and CHIRON spectra for one target (TOI-179) with typical $S/N$ of $\sim$30-40 at around 6000\,\AA.

\section{Analysis and results}
\label{sec:data_analysis}
The analysis of the HARPS-N spectra can be divided into two main steps: the measurement of the stellar atmospheric parameters (effective temperature \teff, surface gravity \logg, and microturbulence velocity \csi), iron and titanium abundances (\feh, \tih), and the measurement of other photospheric elemental abundances fixing the previously derived parameters. Since we performed a differential analysis relative to the Sun, we took as reference the solar parameters and elemental abundances derived by \citet[see that paper for details]{Biazzoetal2022}.

\subsection{Stellar parameters and abundance of iron and titanium}\label{sec:stellar_parameters}

The first step of the analysis was performed following the same approach and same line list as in \cite{baratellaetal2020b}, which exploits both Fe and Ti lines. This procedure is preferred to the standard approach (based only on the Fe lines), especially for the study of young and active stars, like those analysed in this work. The main reason for this preference is that young stars have a high chromospheric variability which affects spectral lines. Since Ti lines form deeper in the photosphere than Fe lines, they are less affected by the chromosphere (see also \citealt{baratellaetal2020a} for details). 

For the measurement of atmospheric parameters, Fe and Ti abundances we employed \textit{pyMOOGi} (\citealt{adamow_2017_pymoogi}) which is a python wrapper for the MOOG code (\citealt{sneden1973}, version 2019). We used the equivalent widths (EWs) method by running the driver \textit{abfind}, which assumes a local thermal equilibrium (LTE). We then considered the 1D model atmospheres linearly interpolated from the ATLAS9 grid of \cite{castelli_kurukz_2003}, with solar-scaled chemical composition and new opacities (ODFNEW). The EWs of the target stars were measured for each line using the ARESv2 code (\citealt{sousa_2015_ares_v2}), but substantial double checks (with the {\it splot} task in IRAF\footnote{IRAF is distributed by the National Optical Astronomy Observatories, which are operated by the Association of Universities for Research in Astronomy, Inc., under a cooperative agreement with the National Science Foundation. NOAO stopped supporting IRAF, see https://iraf-community.github.io/}) were performed, particularly in the blue part of the spectrum because of the intrinsic difficulties in optimal continuum tracing. In this case, depending on the line profile, Gaussian fitting procedure or direct integration was performed. During the analysis, when the EWs were greater than 150\,m\AA\,(strong lines), the related lines were discarded. The same was done for those lines with fitting errors larger than 2$\sigma$. Once the line EWs were measured, \teff\,was derived by imposing the excitation equilibrium, therefore reducing to zero the slope between the abundances of both \ion{Fe}{i} and \ion{Ti}{i} lines and the excitation potential. Then, \ion{Ti}{i} and \ion{Ti}{ii} lines were used to derive \logg\,by imposing the ionisation equilibrium (difference between the mean \ion{Ti}{i} and \ion{Ti}{ii} abundances lower than 0.01\,dex), while \csi\,was obtained by removing the trend between the \ion{Ti}{i} abundances and the reduced equivalent width $REW=\log(EW/\lambda)$ (\citealt{baratellaetal2020b}). We used an iterative procedure by changing parameters at steps of 5\,K in \teff, 0.01\,km/s in \csi, and 0.01\,dex in \logg. The uncertainties on \teff\,and \csi\,were calculated by varying each quantity until the slopes of the relative trends were larger than three times their value with the optimised parameters, while the uncertainty of \logg\,was calculated by varying this parameter until the difference between neutral and ionised Ti became larger than the highest error between the abundances of the two species. We finally derived the stellar parameters with internal accuracy (at 3$\sigma$) ranging from 30 to 100 K in \teff, from 0.05 to 0.17 dex in \logg, from 0.07 to 0.15 km/s in \csi, from 0.08 to 0.12 dex in [Fe/H], and from 0.07 to 0.14\,dex in [Ti/H]. In Table \ref{tab:target_references} we list the final stellar parameters (\teff, \logg, \csi), and the Fe and Ti abundances, which we derived within the GAPS-YOs sub-programme. We note that for seven out of ten targets, the analysis of the stellar parameters was performed within GAPS-YOs for planet validation studies already published or to be submitted. Instead, for the remaining targets, the stellar parameters were derived in a homogeneous way in this work and for the first time within the GAPS project\footnote{We remark that for the two targets TOI-4515 and TOI-179, we considered the stellar parameters derived through only iron lines because the procedure based on Fe+Ti lines was not necessary (see papers cited in Table \ref{tab:target_references} for details).}. Moreover, for the whole sample, no abundance analysis of multiple elements was previously performed.

\subsection{Other elemental abundances through equivalent widths}
\label{sec:abun_anal}

Once the stellar parameters and iron together with titanium abundances were computed, we measured the abundances (\xh\footnote{Throughout the paper the abundance of the X element is reported relative to the solar elemental abundance as \xh\,= log$\bigg(\frac{\epsilon(X)}{\epsilon(H)}\bigg)$ - log$\bigg(\frac{\epsilon(X)}{\epsilon(H)}\bigg)_\odot$, where $\log \epsilon(X)$ is the absolute abundance.}) of the following other 16 elements: lithium (Li), carbon (C), oxygen (O), sodium (Na), magnesium (Mg), aluminium (Al), silicon (Si), sulphur (S), calcium (Ca), chromium (Cr), nickel (Ni), zinc (Zn), yttrium (Y), zirconium (Zr), cerium (Ce)\footnote{Since in \cite{Biazzoetal2022} the cerium abundance for the Sun was not measured, we used the same co-added solar spectrum analysed by the same authors and obtained the following value: $\log \epsilon$(Ce)$_\odot$ = $1.57\pm0.06$.}, and neodymium (Nd), with Cr abundances measured in the first two ionisation states, like Fe and Ti (see Table\,\ref{tab:oth_elem_abund}). For all the 16 elements mentioned above, we adopted the line list in \cite{Biazzoetal2022} and applied the EW method. Taking into account the same grids of model atmospheres, tools, and procedures as those described in Sect.\,\ref{sec:stellar_parameters}, we measured the EWs of each line and discarded the strong lines with EWs greater than 150\,m\AA\,and those with fitting errors larger than 2$\sigma$. 

For those targets observed with the TRES and CHIRON spectrographs, we measured the oxygen abundance through the \ion{O}{i} triplet lines (7771.94\,\AA, 7774.17\,\AA, 7775.39\,\AA) and applied the non-LTE (NLTE) corrections by \cite{amarsi_2015_OI_nlte}.

In addition to \ion{O}{i} triplet lines, we also applied NLTE correction to each line abundance of \ion{C}{i} and \ion{Na}{i} and taking into account the prescriptions by \cite{amarsi_2019_CI_nlte} and \cite{lind_2011_nlte_NaI}, respectively. For the other elements, the corrections were negligible or not reported in the literature (see \citealt{Biazzoetal2022} for details).

We also measured the abundance of lithium by applying the EW procedure to the 6707.8\,\AA\,line. We find that all our targets show the lithium 6707.81\,\AA\,line in their spectra, with the exception of TOI-5543. Table\,\ref{tab:target_references} lists the values of our measurements of lithium equivalent widths and abundances. In the case of TOI-1430 and TOI-4515, we could only derive for $\log \epsilon {\rm (Li)}$ upper limits of 0.1\,dex and 0.5\,dex, respectively. 
Then, we considered the departure from LTE considering the NLTE calculation of \cite{Lind_2009_LiI_nlte}. Final NLTE Li abundances, together with their uncertainties due to both EW measurements and stellar parameters added quadratically, are reported in Table\,\ref{tab:target_references}. A double-check of the final results was done also computing the lithium abundances using the spectral synthesis method (see Sect.\,\ref{sec:spectral_synthesis}).

Finally, we looked for possible trends of final elemental abundances (also for Fe and Ti derived in the previous subsection) with the stellar parameters (\teff, \logg, \csi, \vsini), and with the activity index $\log R'_{\rm HK}$\footnote{As chromospheric activity indicator, we considered the arithmetic average and standard deviation of the \ion{Ca}{ii} H\&K line-core flux, $\log R'_{\rm HK}$.} listed in Table\,\ref{tab:target_references}. The $\log R'_{\rm HK}$ values were measured in this work for each target adopting the colour index B-V extracted from the tables of \cite{pecaut_mamajek_2013_spt} (version 2022) by using our \teff\,values. No relationship was found for our sample.

\setlength{\tabcolsep}{0.99pt}
\begin{table*}
\tiny

\caption{Final stellar parameters, Fe and Ti abundances (the numbers in parenthesis indicate the number of lines used),  $EW_{\rm Li}$ and NLTE lithium abundance, the mean activity index $\log R'_{\rm HK}$, and the stellar rotation period $P_{\rm rot}$ of our sample.
}
\label{tab:target_references}
\begin{center}
\begin{tabular*}{1.009\textwidth}{lrrrrrrrrrrrr}
\hline\hline
\multirow{2}{*}{Name} & \teff & \logg & \csi & [\ion{Fe}{i}/H] & [\ion{Fe}{ii}/H] & [\ion{Ti}{i}/H] & [\ion{Ti}{ii}/H] & $v\sin i$ & $EW_{\rm Li}$ & $\log \epsilon{\rm (Li)}$ & $\log R'_{\rm HK}$ & $P_{\rm rot, obs}$ \\
 & (K) & (dex) & (km/s) & (dex) & (dex) & (dex) & (dex) & (km/s) & (m\AA) & (dex) & (dex) & (days) \\ \hline
TOI-179$^a$ & 5170$\pm$60 & 4.54$\pm$0.09 & 1.12$\pm$0.13 & 0.00$\pm$0.10(80) & 0.00$\pm$0.12(17) & 0.10$\pm$0.06(52) & 0.08$\pm$0.08(25) & 4.8$\pm$0.6 & 40.7$\pm$2.4 & 1.66$\pm$0.07 & $-$4.35$\pm$0.07 & 8.73$\pm$0.07$^a$ \\
TOI-1136$^b$ & 5790$\pm$60 & 4.44$\pm$0.17 & 1.09$\pm$0.13 & 0.03$\pm$0.09(78) & 0.03$\pm$0.09(16) & 0.00$\pm$0.09(50) & 0.00$\pm$0.09(20) & 6.7$\pm$0.6 & 68.3$\pm$1.3 & 2.47$\pm$0.05 & $-$4.41$\pm$0.02 & 8.70$\pm$0.10$^i$ \\
TOI-1430$^c$ & 5075$\pm$50 & 4.55$\pm$0.05 & 0.79$\pm$0.07 & $-$0.02$\pm$0.11(72) & $-$0.01$\pm$0.11(14) & 0.05$\pm$0.10(50) & 0.07$\pm$0.11(21) & 1.9$\pm$0.6 & $<$0.6 & $<$0.10 & $-$4.48$\pm$0.04 & 12.00$\pm$0.40$^b$ \\
TOI-1726$^d$ & 5700$\pm$75 & 4.54$\pm$0.05 & 1.03$\pm$0.10 & 0.03$\pm$0.10(81) & 0.04$\pm$0.13(17) & 0.07$\pm$0.11(44) & 0.09$\pm$0.13(21) & 7.2$\pm$0.7 & 83.5$\pm$2.0 & 2.51$\pm$0.06 & $-$4.39$\pm$0.01 & 6.48$\pm$0.08$^d$ \\
TOI-2048$^b$ & 5100$\pm$100 & 4.59$\pm$0.08 & 0.85$\pm$0.15 & 0.03$\pm$0.11(69) & 0.09$\pm$0.19(14) & 0.13$\pm$0.12(44) & 0.15$\pm$0.14(19) & 8.5$\pm$0.5 & 64.7$\pm$2.0 & 1.83$\pm$0.10 & $-$4.35$\pm$0.02 & 7.97$\pm$0.13$^j$ \\
TOI-2076$^e$ & 5200$\pm$100 & 4.52$\pm$0.05 & 0.90$\pm$0.10 & $-$0.03$\pm$0.10(76) & $-$0.01$\pm$0.11(15) & 0.03$\pm$0.11(47) & 0.06$\pm$0.13(19) & 5.2$\pm$0.4 & 90.2$\pm$2.2 & 2.11$\pm$0.10 & $-$4.40$\pm$0.02 & 7.29$\pm$0.12$^k$ \\
TOI-4515$^f$ & 5450$\pm$30 & 4.48$\pm$0.10 & 1.06$\pm$0.09 & 0.05$\pm$0.09(77) & 0.05$\pm$0.10(16) & 0.00$\pm$0.08(52) & $-$0.03$\pm$0.07(23) & 3.4$\pm$0.5 & $<$2.0 & $<$0.50 & $-$4.47$\pm$0.03 & 15.50$\pm$0.30$^f$ \\
TOI-5082$^b$ & 5680$\pm$65 & 4.52$\pm$0.12 & 1.04$\pm$0.10 & 0.10$\pm$0.08(77) & 0.09$\pm$0.09(16) & 0.10$\pm$0.07(52) & 0.10$\pm$0.09(21) & 6.0$\pm$0.5 & 39.3$\pm$1.5 & 2.10$\pm$0.06 & $-$4.48$\pm$0.04 & 9.00$\pm$1.50$^b$ \\
TOI-5398$^g$ & 6000$\pm$75 & 4.44$\pm$0.10 & 1.12$\pm$0.12 & 0.09$\pm$0.10(85) & 0.08$\pm$0.11(17) & 0.11$\pm$0.10(39) & 0.10$\pm$0.12(24) & 7.5$\pm$0.6 & 89.6$\pm$1.0 & 2.78$\pm$0.06 & $-$4.42$\pm$0.04 & 7.34$\pm$0.05$^g$ \\
TOI-5543$^h$ & 5140$\pm$75 & 4.50$\pm$0.05 & 0.90$\pm$0.15 & 0.11$\pm$0.12(76) & 0.12$\pm$0.17(14) & 0.18$\pm$0.14(56) & 0.20$\pm$0.20(22) & 4.5$\pm$0.5 & ... & ... & $-$4.41$\pm$0.02 & 9.00$\pm$1.00$^b$ \\ 
\hline
\end{tabular*}
\begin{tablenotes}
    \small {
        \item{References:
 $^a$\cite{Desidera_2023_toi179},
 $^b$This Work,
 $^c$Nardiello et al., in prep., 
 $^d$\cite{Damasso_2023_yo40}, $^e$\cite{damasso_2024_toi2076_press}, $^f$\cite{carleoetal2024}, $^g$\cite{Mantovanetal2023}, $^h$Malavolta et al. in prep.,
 $^i$\cite{Dai_2023_so04}, $^j$\cite{Newton_2023_yo42},
 $^k$\cite{nardiello_2022}.
 }
 }
\end{tablenotes}
\end{center}
\end{table*}

\setlength{\tabcolsep}{4pt}
\begin{table*}[ht]
\caption{Final elemental abundances of our stellar sample.}
\label{tab:oth_elem_abund}
\tiny
\begin{center}
\begin{tabular}{lrrrrrrrr}
\hline
\multirow{2}{*}{Name} & [C/H] & [O/H] & [Na/H] & [Mg/H] & [Al/H] & [Si/H] & [S/H] & [Ca/H] \\
 & (dex) & (dex) & (dex) & (dex) & (dex) & (dex) & (dex) & (dex) \\ \hline
TOI-179 & $-$0.19$\pm$0.25(2) & 0.01$\pm$0.09(s,2) & $-$0.04$\pm$0.07(2) & 0.00$\pm$0.10(4) & $-$0.02$\pm$0.07(3) & $-$0.02$\pm$0.09(11) & 0.03$\pm$0.09(2) & 0.04$\pm$0.12(10) \\
TOI-1136 & $-$0.11$\pm$0.06(3) & 0.06$\pm$0.04(s,3) & $-$0.07$\pm$0.02(3) & 0.03$\pm$0.08(4) & $-$0.04$\pm$0.07(3) & 0.03$\pm$0.07(10) & 0.02$\pm$0.08(3) & 0.11$\pm$0.09(11) \\ 
TOI-1430 & $-$0.03$\pm$0.23(2) & 0.15$\pm$0.05(s,3) & $-$0.09$\pm$0.02(2) & $-$0.03$\pm$0.09(4) & $-$0.04$\pm$0.07(2) & $-$0.03$\pm$0.08(10) & 0.11$\pm$0.09(2) & 0.04$\pm$0.10(10) \\
TOI-1726 & $-$0.17$\pm$0.07(3) & 0.01$\pm$0.03(s,3) & $-$0.13$\pm$0.08(4) & 0.01$\pm$0.10(3) & $-$0.06$\pm$0.08(3) & 0.04$\pm$0.09(12) & $-$0.14$\pm$0.09(2) & 0.14$\pm$0.10(13) \\
TOI-2048 & $-$0.03$\pm$0.40(2) & ... & 0.00$\pm$0.08(2) & $-$0.01$\pm$0.10(4) & $-$0.03$\pm$0.08(3) & $-$0.01$\pm$0.09(10) & 0.18$\pm$0.10(2) & 0.11$\pm$0.12(10) \\
TOI-2076 & $-$0.13$\pm$0.24(2) & 0.11$\pm$0.04(s,2) & $-$0.12$\pm$0.04(2) & $-$0.07$\pm$0.12(4) & $-$0.08$\pm$0.06(3) & $-$0.05$\pm$0.08(10) & 0.03$\pm$0.16(2) & 0.05$\pm$0.11(9) \\
TOI-4515 & $-$0.02$\pm$0.09(2) & $-$0.07$\pm$0.05(s,3) & $-$0.09$\pm$0.13(2) & 0.06$\pm$0.11(4) & 0.02$\pm$0.06(2) & 0.04$\pm$0.07(9) & 0.11$\pm$0.09(2) & 0.06$\pm$0.11(10) \\
TOI-5082 & $-$0.07$\pm$0.06(3) & 0.13$\pm$0.07(s,3) & $-$0.02$\pm$0.07(4) & 0.08$\pm$0.08(4) & 0.03$\pm$0.07(3) & 0.12$\pm$0.09(11) & 0.07$\pm$0.08(3) & 0.17$\pm$0.09(11) \\
TOI-5398 & $-$0.05$\pm$0.04(3) & 0.16$\pm$0.04(s,3) & $-$0.01$\pm$0.03(4) & 0.01$\pm$0.08(4) & $-$0.05$\pm$0.07(2) & 0.06$\pm$0.10(12) & 0.05$\pm$0.21(2) & 0.16$\pm$0.08(12) \\
TOI-5543 & 0.12$\pm$0.18(2) & 0.10$\pm$0.22(s) & 0.18$\pm$0.10(2) & 0.09$\pm$0.12(4) & 0.08$\pm$0.07(3) & 0.14$\pm$0.12(10) & 0.21$\pm$0.09(3) & 0.13$\pm$0.12(9) \\
\hline
\multirow{2}{*}{Name} & [\ion{Cr}{i}/H] & [\ion{Cr}{ii}/H] & [Ni/H] & [Zn/H] & [Y/H] & [Zr/H] & [Ce/H] & [Nd/H] \\
 & (dex) & (dex) & (dex) & (dex) & (dex) & (dex) & (dex) & (dex) \\ \hline
TOI-179 & 0.12$\pm$0.10(10) & 0.05$\pm$0.05(1) & $-$0.08$\pm$0.08(41) & $-$0.14$\pm$0.08(3) & 0.15$\pm$0.20(5) & 0.08$\pm$0.09(3) & 0.23$\pm$0.07(4) & 0.25$\pm$0.11(2) \\
TOI-1136 & 0.09$\pm$0.06(15) & 0.09$\pm$0.08(6) & $-$0.02$\pm$0.07(45) & 0.06$\pm$0.07(2) & 0.16$\pm$0.08(6) & 0.12$\pm$0.05(3) & 0.25$\pm$0.09(3) & 0.37$\pm$0.09(3) \\ 
TOI-1430 & 0.09$\pm$0.07(12) & 0.10$\pm$0.08(4) & $-$0.06$\pm$0.08(41) & $-$0.03$\pm$0.07(2) & 0.11$\pm$0.13(5) & 0.24$\pm$0.12(3) & 0.20$\pm$0.09(5) & 0.13$\pm$0.04(2) \\
TOI-1726 & 0.12$\pm$0.06(15) & 0.10$\pm$0.06(6) & $-$0.03$\pm$0.06(43) & 0.06$\pm$0.06(2) & 0.17$\pm$0.10(5) & 0.19$\pm$0.05(3) & 0.19$\pm$0.10(4) & 0.31$\pm$0.02(2) \\
TOI-2048 & 0.21$\pm$0.08(12) & 0.17$\pm$0.18(2) & 0.00$\pm$0.08(38) & 0.09$\pm$0.09(3) & 0.14$\pm$0.32(2) & 0.21$\pm$0.08(3) & 0.32$\pm$0.07(4) & 0.02$\pm$0.11(2) \\
TOI-2076 & 0.09$\pm$0.07(15) & 0.09$\pm$0.10(6) & $-$0.09$\pm$0.07(39) & $-$0.01$\pm$0.09(3) & 0.13$\pm$0.13(5) & 0.07$\pm$0.05(3) & 0.10$\pm$0.11(4) & $-$0.05$\pm$0.15(2) \\
TOI-4515 & 0.12$\pm$0.06(13) & 0.05$\pm$0.09(6) & 0.01$\pm$0.08(42) & $-$0.09$\pm$0.06(2) & 0.06$\pm$0.12(5) & $-$0.05$\pm$0.02(2) & 0.10$\pm$0.07(5) & 0.07$\pm$0.06(3) \\
TOI-5082 & 0.19$\pm$0.06(13) & 0.18$\pm$0.08(6) & 0.09$\pm$0.08(45) & 0.14$\pm$0.09(2) & 0.26$\pm$0.10(4) & 0.20$\pm$0.05(3) & 0.31$\pm$0.09(4) & 0.36$\pm$0.04(2) \\
TOI-5398 & 0.11$\pm$0.09(15) & 0.10$\pm$0.08(6) & 0.01$\pm$0.06(38) & 0.10$\pm$0.09(2) & 0.22$\pm$0.10(5) & 0.17$\pm$0.06(3) & 0.23$\pm$0.09(4) & 0.33$\pm$0.07(3) \\
TOI-5543 & 0.20$\pm$0.07(11) & 0.20$\pm$0.12(4) & 0.11$\pm$0.09(40) & 0.15$\pm$0.11(3) & 0.20$\pm$0.14(4) & 0.12$\pm$0.06(3) & 0.22$\pm$0.09(4) & 0.08$\pm$0.02(2) \\
\hline
\end{tabular}
\end{center}
\begin{tablenotes}
    \small {
        \item{Note:
The number of lines used in the analysis and the cases for which we applied the spectral synthesis method (s) are shown in parentheses.
 }
 }
\end{tablenotes}
\end{table*}

\subsection{Rotational velocities and elemental abundances through spectral synthesis}
\label{sec:spectral_synthesis}

The projected rotational velocity (\vsini) was measured through the spectral synthesis technique and considering the {\it synth} driver of the same codes used in Sect.\,\ref{sec:stellar_parameters}. We then created synthetic spectra from the \cite{castelli_kurukz_2003} grids of model atmospheres fixing the stellar parameters \teff, \logg, \csi, and \feh\,to those derived in Sect.\,\ref{sec:stellar_parameters}. We convolved the resulting synthetic spectra with a Gaussian profile corresponding to the resolving power of HARPS-N of $R\sim115\,000$, taking into account the optical limb-darkening coefficients by \cite{Claret_2019} at the \teff, \logg, \csi, and \feh\, of our targets. As done in \cite{Biazzoetal2022}, we applied the empirical relationships for the macroturbulence velocity ($V_{\rm macro}$) by \cite{doyle_2014} for stellar $T_{\rm eff}$>5700 K, while empirical relationships by \cite{Brewer_2016} were considered for $T_{\rm eff}$<5700 K\,(see that paper for details). Finally, we synthesised spectral lines in three regions around 5400, 6200, and 6700\,\AA, as done in, for example, \cite{carleoetal2024}, until the minimum of residuals between stellar and synthetic spectra were reached. The final values are listed in Table\,\ref{tab:target_references}, where uncertainties in stellar parameters together with spectral continuum definition were taken into account for the \vsini\,error estimates (for the uncertainties in the continuum position, see Sect.\,\ref{sec:uncertainty_EW_synth}).

We also applied the same spectral synthesis method to derive the oxygen abundance using the forbidden [\ion{O}{i}] line at 6300.3\,\AA. Atomic parameters for the oxygen line and the nearby (blended) \ion{Ni}{i} line, which exhibits an isotopic structure, were taken from \cite{Caffau_2008} and \cite{johansson_2003}, respectively. Since nickel abundances were needed for the synthesis of the oxygen line profile, their values were fixed to the ones we derived above (see Sect.\,\ref{sec:abun_anal}). Since the 6300.3\AA\,line is not affected by deviation from LTE, we did not apply any NLTE corrections (see \citealt{Caffau_2008} for details). In the end, when TRES and CHIRON spectra were available, O abundances were derived with both EWs and spectral synthesis methods, therefore as final abundances, we considered the weighted average coming from these two methods. 

Finally, the spectral synthesis technique was also adopted for the lithium abundance ($\log \epsilon{\rm (Li)}$) measurements. Therefore, we used the line list by \cite{Reddyetal2002} in the vicinity of the Li\,6707.8\,\AA\,line, implemented with the VALD database (\citealt{kupkaetal2000}) for wavelengths farther from the line centre, as done in \cite{Biazzoetal2022}. The lithium abundance derived through the same codes and model atmospheres used in Sect.\,\ref{sec:stellar_parameters} was then corrected for the departure from LTE considering the NLTE calculations of \cite{Lind_2009_LiI_nlte}. We found consistent values with both methods based on line EW and synthesis within the uncertainties. In Table\,\ref{tab:target_references} we list only the results obtained through the EW method, with the exception of TOI-1430 and TOI-4515 for which only the synthesis method allowed us to derive upper limits in $\log \epsilon{\rm (Li)}$.

\subsection{Uncertainties in elemental abundances}
\label{sec:uncertainty_EW_synth}

The errors in the abundances \xh\,are mainly due to uncertainties in atomic parameters, stellar parameters, and EW measurements or continuum position (for the spectral synthesis method). 

Uncertainties in atomic parameters should be negligible since our analysis is performed differentially with respect to the Sun. 

Errors due to stellar parameter uncertainties were estimated by varying each parameter separately by its assessed error while keeping the others unchanged. As shown in Table\,\ref{tab:target_references}, the uncertainties in effective temperature are in the range 30-100\,K (with standard deviation $\sigma$ = 20\,K), 0.05-0.17\,dex for $\log g$ (with $\sigma$ = 0.04\,dex), 0.07-0.15\,km/s for $\xi$ (with $\sigma$ = 0.03\,km/s), 0.08-0.12\,dex for [Fe/H] (with $\sigma$ = 0.01\,dex), and 0.07-0.14\,dex for [Ti/H] ($\sigma$ = 0.02\,dex). Due to the small values of the standard deviations and the relatively small range in all stellar parameters and Fe and Ti abundances (among the others with very similar $\sigma$), we decided to consider typical uncertainties in stellar parameters of 70\,K, 0.10\,dex, 0.10\,km/s, 0.10\,dex in \teff, \logg, $\xi$, and \feh, respectively. Errors in elemental abundances [X/H] due to uncertainties in individual stellar parameters are listed in Table\,\ref{tab:errors} for, respectively, the coolest and the warmest targets in our sample.

The errors in $\log \epsilon$(X) due to uncertainties in EWs were obtained by computing the standard deviation around the mean abundance determined from all the measured lines of each element. From these values, the uncertainties in [X/H] were obtained by the quadratic sum of the error for the target and the error for the Sun, as shown in Tables\,\ref{tab:target_references} and \ref{tab:oth_elem_abund} together with the number of lines employed for the abundance analysis (in brackets). When only one line was measured (such as for the lithium element), the error we used was the standard deviation of three independent EW measurements obtained considering three different positions of the continuum.

The uncertainties on the oxygen abundances derived through the spectral synthesis method are of two types: $i.$ errors related to the best-fit determination, which were evaluated by changing the continuum position until the standard deviation was two times larger than the best-fit value; typical uncertainties, in this case, are around 0.12\,dex; $ii.$ errors related to the uncertainty in the nickel abundance measurements, as the oxygen line at 6300.3\,\AA\, is blended with a Ni line (see Sect.\,\ref{sec:spectral_synthesis}; typical uncertainties due to Ni are around 0.05\,dex). These sources of internal errors were quadratically added to obtain the total error (see Table\,\ref{tab:oth_elem_abund}).

\setlength{\tabcolsep}{6pt}
\begin{table*}
\begin{center}
\caption{Internal errors on abundance measurements due to uncertainties in stellar parameters.
}
\label{tab:errors}
\small
\begin{tabular}{lcccc}
\hline
\multirow{2}{*}{TOI-1430} & $T_{\rm eff}=5075$ K & $\log g=4.55$\,dex & $\xi=0.79$ km/s & \feh=$-0.02$\,dex \\ \cline{2-5} 
 & $\Delta T_{\rm eff}=-/+70$ K & $\Delta \log g=-/+0.10$\,dex & $\Delta \xi=-/+0.10$ km/s & $\Delta$[Fe/H]$=-/+0.10$\,dex \\ \hline
$[$\ion{Fe}{i}/H$]$ & $-$0.04/0.05 & 0.01/$-$0.01 & 0.03/$-$0.02 & .../... \\
$[$\ion{Fe}{ii}/H$]$ & 0.07/0.01 & 0.01/0.09 & 0.06/0.02 & .../... \\
$[$\ion{C}{i}/H$]$ & 0.05/$-$0.06 & $-$0.04/0.06 & 0.00/0.00 & 0.01/0.01 \\
$[$\ion{O}{i}/H$]$ & 0.08/$-$0.09 & $-$0.05/0.08 & 0.00/0.00 & $-$0.03/0.05 \\
$[$\ion{Na}{i}/H$]$ & $-$0.05/0.05 & 0.02/$-$0.01 & 0.01/$-$0.01 & $-$0.01/0.01 \\
$[$\ion{Mg}{i}/H$]$ & $-$0.02/0.03 & 0.02/$-$0.01 & 0.01/0.00 & $-$0.02/0.03 \\
$[$\ion{Al}{i}/H$]$ & $-$0.04/0.04 & 0.01/$-$0.01 & 0.01/$-$0.01 & $-$0.01/0.01 \\
$[$\ion{Si}{i}/H$]$ & 0.03/$-$0.02 & $-$0.01/0.02 & 0.00/0.00 & $-$0.02/0.03 \\
$[$\ion{S}{i}/H$]$ & 0.04/$-$0.05 & $-$0.04/0.05 & 0.00/0.00 & 0.01/0.01 \\
$[$\ion{Ca}{i}/H$]$ & $-$0.06/0.07 & 0.04/$-$0.04 & 0.02/$-$0.01 & $-$0.02/0.02 \\
$[$\ion{Ti}{i}/H$]$ & $-$0.07/0.06 & 0.00/$-$0.02 & 0.02/$-$0.02 & 0.00/$-$0.01 \\
$[$\ion{Ti}{ii}/H$]$ & 0.04/0.02 & 0.00/0.08 & 0.06/0.01 & 0.00/0.07 \\
$[$\ion{Cr}{i}/H$]$ & $-$0.06/0.07 & 0.02/$-$0.02 & 0.02/$-$0.02 & $-$0.02/0.02 \\
$[$\ion{Cr}{ii}/H$]$ & 0.04/$-$0.04 & $-$0.04/0.06 & 0.01/$-$0.01 & $-$0.02/0.04 \\
$[$\ion{Ni}{i}/H$]$ & 0.00/0.01 & $-$0.01/0.02 & 0.02/$-$0.01 & $-$0.03/0.04 \\
$[$\ion{Zn}{i}/H$]$ & 0.02/$-$0.01 & 0.00/0.02 & 0.03/$-$0.02 & $-$0.03/0.05 \\
$[$\ion{Y}{ii}/H$]$ & 0.00/0.01 & $-$0.03/0.04 & 0.04/$-$0.03 & $-$0.03/0.05 \\
$[$\ion{Zr}{ii}/H$]$ & 0.00/0.01 & $-$0.04/0.04 & 0.03/$-$0.02 & $-$0.04/0.05 \\
$[$\ion{Ce}{i}/H$]$ & $-$0.01/0.01 & $-$0.04/0.05 & 0.01/0.00 & $-$0.03/0.04 \\
$[$\ion{Nd}{ii}/H$]$ & $-$0.01/0.02 & $-$0.04/0.05 & 0.01/0.00 & $-$0.04/0.04 \\ \hline
\multirow{2}{*}{TOI-5398} & $T_{\rm eff}=6000$ K & $\log g=4.44$\,dex & $\xi=1.12$ km/s & \feh=0.09\,dex \\ \cline{2-5} 
 & $\Delta T_{\rm eff}=-/+70$ K & $\Delta \log g=-/+0.10$\,dex & $\Delta \xi=-/+0.10$ km/s & $\Delta$[Fe/H]$=-/+0.10$\,dex \\ \hline
$[$\ion{Fe}{i}/H$]$ & $-$0.05/0.05 & 0.00/$-$0.01 & 0.02/$-$0.02 & .../... \\
$[$\ion{Fe}{ii}/H$]$ & 0.03/$-$0.01 & $-$0.03/0.04 & 0.03/$-$0.02 & .../... \\
$[$\ion{C}{i}/H$]$ & 0.05/$-$0.03 & $-$0.03/0.03 & 0.00/0.00 & 0.00/0.00 \\
$[$\ion{O}{i}/H$]$ & 0.07/$-$0.04 & $-$0.04/0.04 & 0.01/$-$0.01 & $-$0.04/0.03 \\
$[$\ion{Na}{i}/H$]$ & $-$0.03/0.04 & 0.01/$-$0.02 & 0.01/$-$0.01 & 0.00/0.00 \\
$[$\ion{Mg}{i}/H$]$ & $-$0.02/0.03 & 0.01/$-$0.01 & 0.01/$-$0.01 & 0.00/0.00 \\
$[$\ion{Al}{i}/H$]$ & $-$0.02/0.03 & 0.00/0.00 & 0.00/0.00 & 0.00/0.00 \\
$[$\ion{Si}{i}/H$]$ & $-$0.01/0.02 & 0.00/0.00 & 0.01/0.00 & 0.00/0.01 \\
$[$\ion{S}{i}/H$]$ & 0.04/$-$0.02 & $-$0.02/0.03 & 0.00/0.00 & 0.00/0.00 \\
$[$\ion{Ca}{i}/H$]$ & $-$0.04/0.05 & 0.02/$-$0.02 & 0.02/$-$0.02 & 0.00/0.00 \\
$[$\ion{Ti}{i}/H$]$ & $-$0.06/0.07 & 0.00/0.00 & 0.02/$-$0.01 & 0.00/0.00 \\
$[$\ion{Ti}{ii}/H$]$ & 0.01/0.01 & $-$0.02/0.05 & 0.04/$-$0.01 & $-$0.02/0.04 \\
$[$\ion{Cr}{i}/H$]$ & $-$0.05/0.05 & 0.00/0.00 & 0.02/$-$0.02 & 0.00/0.00 \\
$[$\ion{Cr}{ii}/H$]$ & 0.03/$-$0.01 & $-$0.03/0.04 & 0.02/$-$0.02 & $-$0.02/0.02 \\
$[$\ion{Ni}{i}/H$]$ & $-$0.03/0.04 & 0.00/0.00 & 0.01/$-$0.01 & 0.00/0.00 \\
$[$\ion{Zn}{i}/H$]$ & $-$0.01/0.02 & $-$0.01/0.00 & 0.05/$-$0.05 & $-$0.02/0.02 \\
$[$\ion{Y}{ii}/H$]$ & 0.00/0.01 & $-$0.04/0.04 & 0.04/$-$0.04 & $-$0.03/0.03 \\
$[$\ion{Zr}{ii}/H$]$ & 0.00/0.01 & $-$0.04/0.04 & 0.02/$-$0.02 & $-$0.03/0.03 \\
$[$\ion{Ce}{i}/H$]$ & $-$0.01/0.02 & $-$0.04/0.04 & 0.01/$-$0.01 & $-$0.03/0.03 \\
$[$\ion{Nd}{ii}/H$]$ & $-$0.01/0.02 & $-$0.04/0.04 & 0.00/$-$0.01 & $-$0.03/0.03 \\ \hline
\end{tabular}
\end{center}
Notes. Measurements were done for TOI-1430 (for all elements but \ion{Fe}{i}, \ion{Fe}{ii}, \ion{Ti}{i}\,and \ion{Ti}{ii} which refer to TOI-5082) and TOI-5398 (for all elements but \ion{Fe}{i}, \ion{Fe}{ii}, \ion{Ti}{i}\,and \ion{Ti}{ii} which refer to TOI-1136), respectively the coolest and the warmest stars in our sample. The values in table refer to the differences between the abundances obtained with ($-$ and $+$) and without the uncertainties in stellar parameters. 
\end{table*}


\section{Discussion}
\label{sec:result_discussion}

\subsection{Elemental abundances in the context of the Galactic disk}

\begin{figure*}[t]
    \centering    \includegraphics[width=\textwidth]{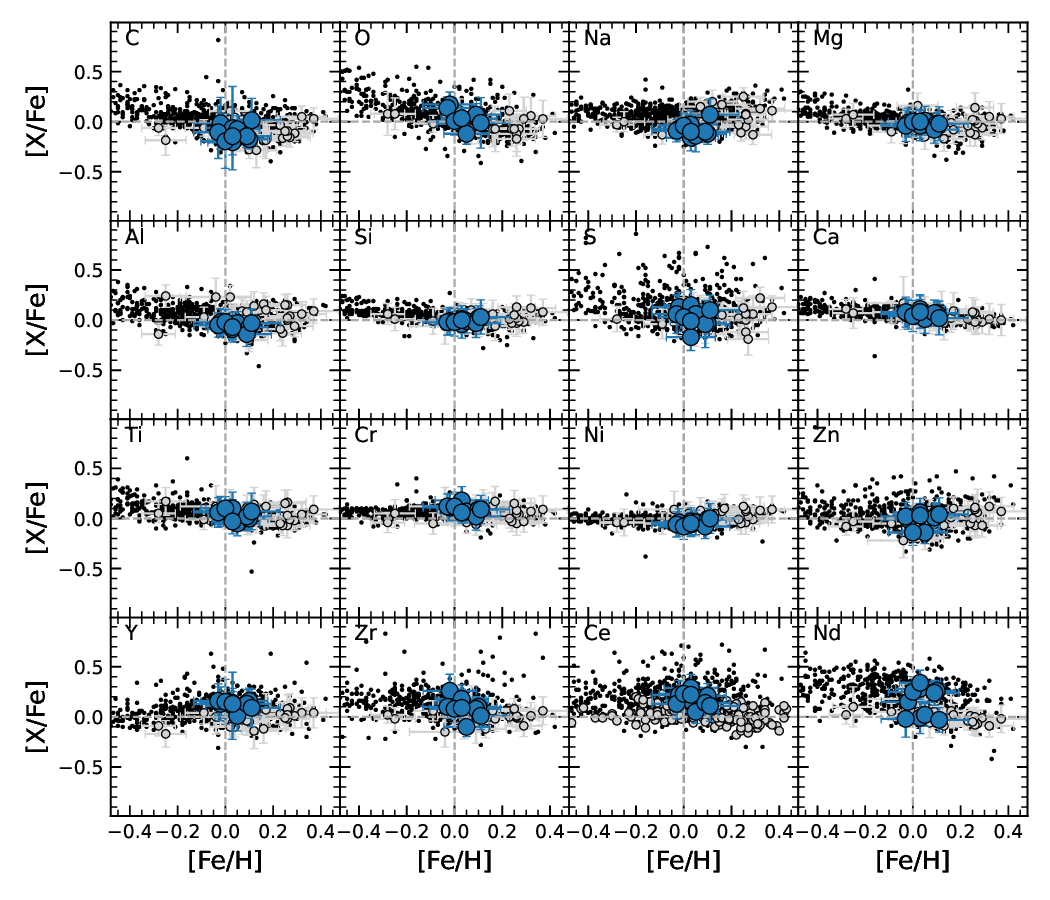}
    \caption{[X/Fe] versus [Fe/H] for our sample (filled blue dots). Horizontal and vertical dashed lines mark the solar abundances. Filled grey dots represent the [X/Fe] values of old transiting planet host stars analysed by \cite{Biazzoetal2022}, except for Ce for which we plot the values by \cite{DelgadoMenaetal2017}. Small black dots in the background represent the [X/Fe] distribution of nearby field stars (distance $\leq$ 200 pc) in the Galactic thin disk with similar parameters as those of our sample (\citealt{Hinkeletal2014}). 
    }
    \label{fig:_X_Fe_vs_Fe_H_}
\end{figure*}

Each Galactic component (thin and thick disk, bulge, and halo) is characterised by different chemical abundance patterns when compared to the iron abundance. The observed differences in these abundance patterns are the consequence of a variety of star formation histories. Therefore, possible relationships between elemental abundances and Fe abundances are generally used to trace the chemical evolution of our Galaxy (\citealt{Bensbyetal2014}). In particular, the study of low-mass stars can be very useful in explaining the history of the Galactic chemical evolution.
The chemistry of a star is indeed intimately connected to the time and place of its birth (see \citealt{adibek2011} and references therein). In order to check that this is indeed true also for our sample of relatively young transiting exoplanet host stars, we compare [X/Fe]\footnote{Throughout the paper the elemental abundance ratios of the X element with Fe is given as [X/Fe] = \xh\,- [Fe/H]} ratio versus [Fe/H] of our targets with abundances of nearby (distance $<$200 pc) field stars of the Galactic thin disk selected from the Hypatia Catalogue (\citealt{Hinkeletal2014}). In particular, we considered targets with stellar parameters similar to those of our sample (i.e. $5000 < T_{\rm eff} \, {(\rm K)} < 6000$, $4.4 <\log g \, {(\rm dex)} < 4.6$). Figure\,\ref{fig:_X_Fe_vs_Fe_H_} shows overall good agreement between the [X/Fe] versus [Fe/H] distribution of our targets and the nearby field stars with iron abundance close to zero. In the same figure, we overplot the abundances homogeneously derived within GAPS (except for Ce, taken from \citealt{DelgadoMenaetal2017}) for older FGK transiting planet-hosting stars. The agreement is clear, with our targets compatible with solar values. 

\begin{figure}[h]
    \centering
    \includegraphics[width = \linewidth]{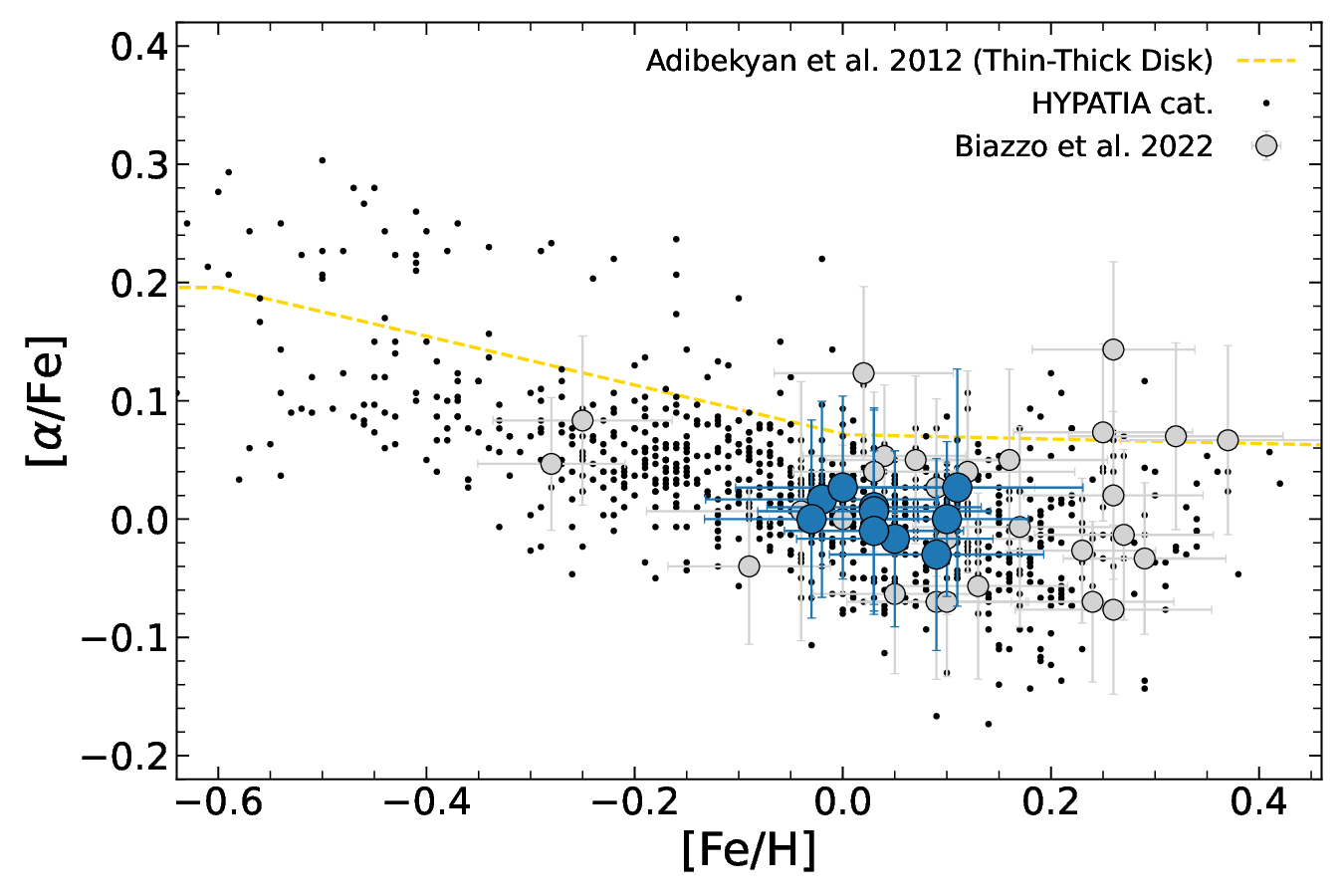}
    \caption{[$\alpha$/Fe] versus [Fe/H]. The overplotted grey points are targets from \cite{Biazzoetal2022}. The dashed yellow line is the fiducial division into thin- and thick-field disk populations, as defined by \cite{adibek2012c}. Small black dots in the background represent the [$\alpha$/Fe] distribution of nearby field stars representative of the nearby Galactic disk taken from the Hypatia Catalogue (\citealt{Hinkeletal2014}).}
    \label{fig:alphafe_feh}
\end{figure}

The thin- and thick-disk stars can also be chemically distinguished by looking at their $\alpha$-element content at a given iron abundance. \cite{adibek2011} showed that for dwarf stars the two populations are clearly separated in terms of [$\alpha$/Fe]\footnote{The $\alpha$ index was estimated as the average of the abundances of Mg, Si, Ti: [$\alpha$/Fe]=$\frac{1}{3}$[Mg/Fe]+[Si/Fe]+[Ti/Fe], as in \cite{adibek2012c} and \cite{Biazzoetal2022}.} from low metallicities up to super-solar metallicities. The [$\alpha$/Fe] versus [Fe/H] position obtained for our targets is depicted in Fig.\ref{fig:alphafe_feh}, together with the division by \cite{adibek2012c} separating thin- and thick- disk populations (yellow dashed line). The corresponding values taken from the Hypatia Catalogue for nearby dwarf stars with similar parameters as our targets are overplotted with small black dots (\citealt{Hinkeletal2014}). It is evident that our targets (filled blue dots) are all below the dashed line and around solar [$\alpha$/Fe] (and [Fe/H]) values, consistently with the position of Galactic thin stars.

\subsection{Kinematic properties in the context of the Galactic disk}
\label{sec:kinematic_prop}

\begin{figure*} 
\begin{center}
\includegraphics[width=8cm,angle=90]{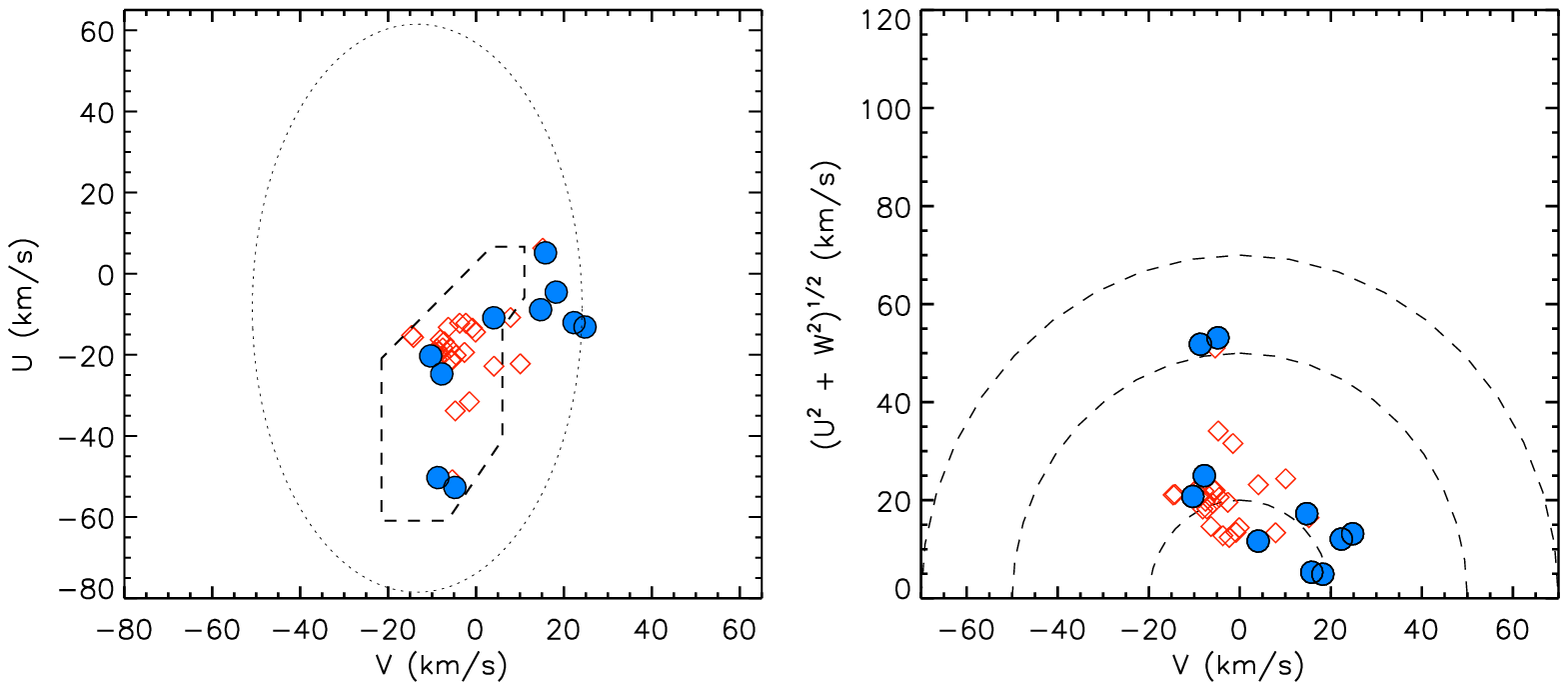}
\caption{Position of our planet-hosting stars in the Boettlinger and Toomre diagrams, where errors are within the symbol dimension. Open diamonds in both panels represent the position of young and nearby clusters analysed by \cite{gagneetal2018}. Left panel: The dashed line represents the boundary separating young disk and old disk stars in the $(U, V)$ plane according to \cite{eggen1996}, while the dotted line shows the velocity ellipsoid defined by \cite{FrancisAnderson2009} for young stars. Right panel: The dashed lines indicate constant peculiar total velocities $v_{\rm tot}=(U^2+V^2+W^2)^{1/2}$= 20, 50, and 70 km/s.}
\label{fig:space_motion} 
\end{center}
\end{figure*}

After the chemical analysis performed in the previous sections, we now characterise the population of our targets in terms of their kinematics. We expect indeed that not only the chemistry, but also the dynamical history of stars, including stellar kinematics, may impact the distribution and the architecture of planets (see \citealt{Magrinietal2022, Biazzoetal2022}, and references therein). Using the mean radial velocities ($<V_{\rm rad}>$) from the HARPS-N spectra, parallaxes ($\pi$) and proper motions ($\mu_{\alpha}, \mu_{\delta}$) from {\it Gaia}\,DR3 (\citealt{gaiacollaborationetal2016, gaiacollaborationetal2021}), we computed space velocity components $UVW$ with respect to the local standard of rest (LSR), correcting for the solar motion derived by \cite{coskunogluetal2011}:  ($U_\odot$, $V_\odot$, $W_\odot$)=(8.50, 13.38, 6.49) km/s. We then considered the general outline of \cite{johnsonsoderblom1987} in a right-hand coordinate system (i.e. with $U$ pointing towards the Galactic centre, $V$ towards the local direction of rotation in the plane of the Galaxy, and $W$ towards the North Galactic Pole). The uncertainties in the space velocity components were obtained considering the prescriptions by \cite{gagneetal2014} and taking into account the error contributions of $V_{\rm rad}$, $\pi$, $\mu_{\alpha}$, and $\mu_{\delta}$. Combining the errors on parallaxes, proper motions, and radial velocities, the resulting average uncertainties in the $U$, $V$, $W$ velocities are around 0.15\,km/s (for more detail see Table\,\ref{tab:space_motion}). In the left panel of Fig.~\ref{fig:space_motion}, we show the location of our targets in the Boettlinger diagram in the ($U$, $V$) plane, where the boundary containing nearby ($<$150\,pc) associations younger than $\sim 1$\,Gyr is shown with a dashed line (see \citealt{eggen1996, gagneetal2018}). The locus containing young stars as defined by \cite{FrancisAnderson2009} is also overplotted (dotted line). All our targets are close to or within the young and nearby Galactic thin disk boundary and all of them are within the locus defined by \cite{FrancisAnderson2009}. In the right panel of Fig.\,\ref{fig:space_motion} we plot the position of our targets in the Toomre diagram, which represents the quadratic addition of the radial and vertical kinetic energies ($v_{\rm tot}$) versus the rotational component. In first approximation, stars with a total velocity lower than 50\,km/s belong to the Galactic thin disk, while those with $70 \ltsim v_{\rm tot} \ltsim 180$\,km/s are likely to be thick disk stars (\citealt{Nissen2004, Bensbyetal2014}). As expected, all our targets are part of the thin disk population, with $v_{\rm tot}$ within or nearby the limit of $\sim 50$\,km/s, while some of them, with $v_{\rm tot}$ around 20\,km/s, are very close to the Sun.

\setlength{\tabcolsep}{4.3pt}
\begin{table*}[ht!]
\caption{Basic information and stellar kinematic properties of our targets. }
\label{tab:space_motion}
\small
\begin{tabular}{lrrrcrrrrrc}
\hline
\multirow{2}{*}{Name} & $\alpha$ & $\delta$ & Parallax & \multirow{2}{*}{Sp.T.} & $G$ & $U$ & $V$ & $W$ & \multirow{2}{*}{$TD/D$} & \multirow{2}{*}{Group} \\
 & (deg) & (deg) & (mas) &  & (mag) & (km/s) & (km/s) & (km/s) &  &  \\ \hline
TOI-179 & 44.26$\pm$0.01 & $-$56.19$\pm$0.01 & 25.88$\pm$0.01 & K1V & 8.74 & $-$12.06$\pm$0.05 & 22.31$\pm$0.72 & 0.26$\pm$0.96 & 0.02 & Field$^a$ \\
TOI-1136 & 192.18$\pm$0.01 & 64.86$\pm$0.01 & 11.82$\pm$0.01 & G2V & 9.38 & $-$8.87$\pm$0.07 & 14.71$\pm$0.10 & 14.78$\pm$0.16 & 0.02 & Field$^b$ \\ 
TOI-1430 & 300.61$\pm$0.01 & 53.38$\pm$0.01 & 24.25$\pm$0.01 & K2V & 8.95 & $-$50.27$\pm$0.02 & $-$8.65$\pm$0.20 & $-$12.53$\pm$0.04 & 0.02 & Field$^c$ \\
TOI-1726 & 117.48$\pm$0.02 & 27.36$\pm$0.01 & 44.68$\pm$0.02 & G3V & 6.74 & 5.14$\pm$0.18 & 15.81$\pm$0.04 & $-$1.27$\pm$0.08 & 0.01 & Ursa Major$^d$ \\
TOI-2048 & 237.92$\pm$0.01 & 52.31$\pm$0.01 & 8.59$\pm$0.01 & K2V & 11.32 & $-$10.87$\pm$0.02 & 4.04$\pm$0.13 & 4.20$\pm$0.15 & 0.01 & \footnotesize Group X$^e$ \\
\multirow{2}{*}{TOI-2076} & \multirow{2}{*}{217.39$\pm$0.01} & \multirow{2}{*}{39.79$\pm$0.01} & \multirow{2}{*}{23.81$\pm$0.01} & \multirow{2}{*}{K1V} & \multirow{2}{*}{8.92} & \multirow{2}{*}{$-$24.71$\pm$0.03} & \multirow{2}{*}{$-$7.78$\pm$0.08} & \multirow{2}{*}{3.76$\pm$0.18} & \multirow{2}{*}{0.01} & \scriptsize Comoving pair \\
 &  &  &  &  &  &  &  &  &  & \scriptsize with TOI-1807$^f$ \\
TOI-4515 & 21.19$\pm$0.02 & 21.51$\pm$0.01 & 5.16$\pm$0.02 & G8V & 11.80 & $-$13.14$\pm$0.11 & 24.79$\pm$0.11 & $-$0.14$\pm$0.13 & 0.02 & Field$^g$ \\
TOI-5082 & 106.57$\pm$0.02 & 22.68$\pm$0.02 & 23.24$\pm$0.03 & G5V & 8.09 & $-$52.69$\pm$0.19 & $-$4.80$\pm$0.05 & $-$6.80$\pm$0.05 & 0.02 & $\dots$ \\
TOI-5398 & 161.88$\pm$0.01 & 36.33$\pm$0.01 & 7.62$\pm$0.01 & F9.5V & 9.99 & $-$4.55$\pm$0.09 & 18.25$\pm$0.01 & $-$1.83$\pm$0.18 & 0.01 & Field$^h$ \\
TOI-5543 & 66.14$\pm$0.02 & 23.27$\pm$0.01 & 7.99$\pm$0.02 & K2V & 11.23 & $-$20.28$\pm$0.19 & $-$10.31$\pm$0.06 & $-$4.45$\pm$0.08 & 0.01 & $\dots$ \\
\hline
\end{tabular}
\begin{tablenotes}
    \small {
 \item{Notes: columns list right ascensions, declinations, \textit{Gaia} DR3 magnitudes, and parallaxes taken from \cite{gaiacollaborationetal2016, gaiacollaborationetal2021}, spectral types obtained from the \cite{pecaut_mamajek_2013_spt} table (version 2022) and our \teff, $UVW$ velocities, and thick-to-thin disk probability (TD/D) as derived in the present work. The last column lists the membership group from the literature.}      \item{References: 
 $^a$\cite{Desidera_2023_toi179}, $^b$\cite{Dai_2023_so04},
 $^c$Nardiello et al. in prep., $^d$\cite{Mannetal2020}, $^e$\cite{Newton_2023_yo42}, $^f$\cite{hedges_2021_toi2076_yo43}, $^g$\cite{carleoetal2024}, $^h$\cite{Mantovanetal2023}.}
    }
\end{tablenotes}
\end{table*}

Finally, we also computed the thick-to-thin disk probability ratio, considering the prescriptions reported by \cite{Bensbyetal2014} for the Gaussian distribution of random velocities of different stellar populations. In particular, to get the probabilities $D$ and $TD$ for the thin and thick disk membership, we considered the asymmetric drift, the velocity dispersion, and the fraction of each population given by \cite{Bensbyetal2014}. Our targets show TD/D ratio of $\sim 0.01-0.02$, which is compatible with the membership to the Galactic thin disk (Table\,\ref{tab:space_motion}).

\subsection{Lithium as a chemical clock and comparison with gyrochronological and activity dating indicators}
\label{sec:ages}

\begin{figure*}[t]
    \centering
    \includegraphics[width=0.48\textwidth]{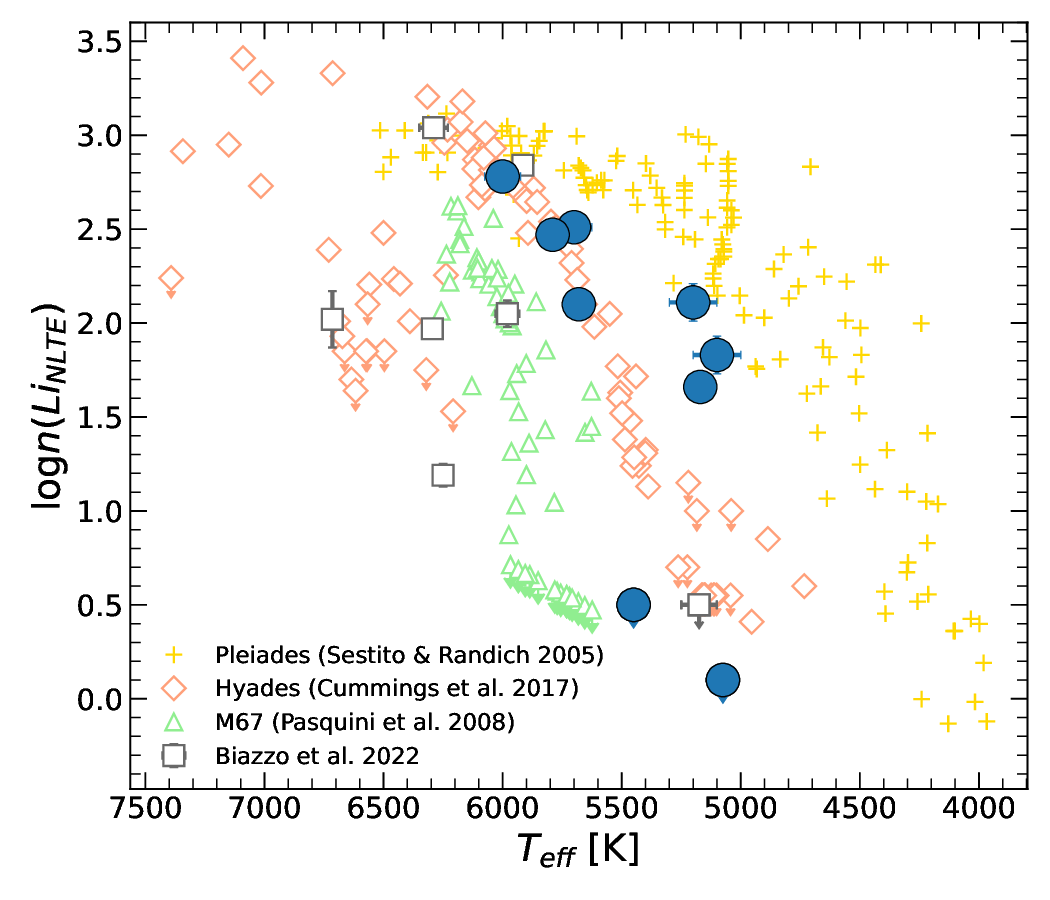}
    \includegraphics[width=0.48\textwidth]{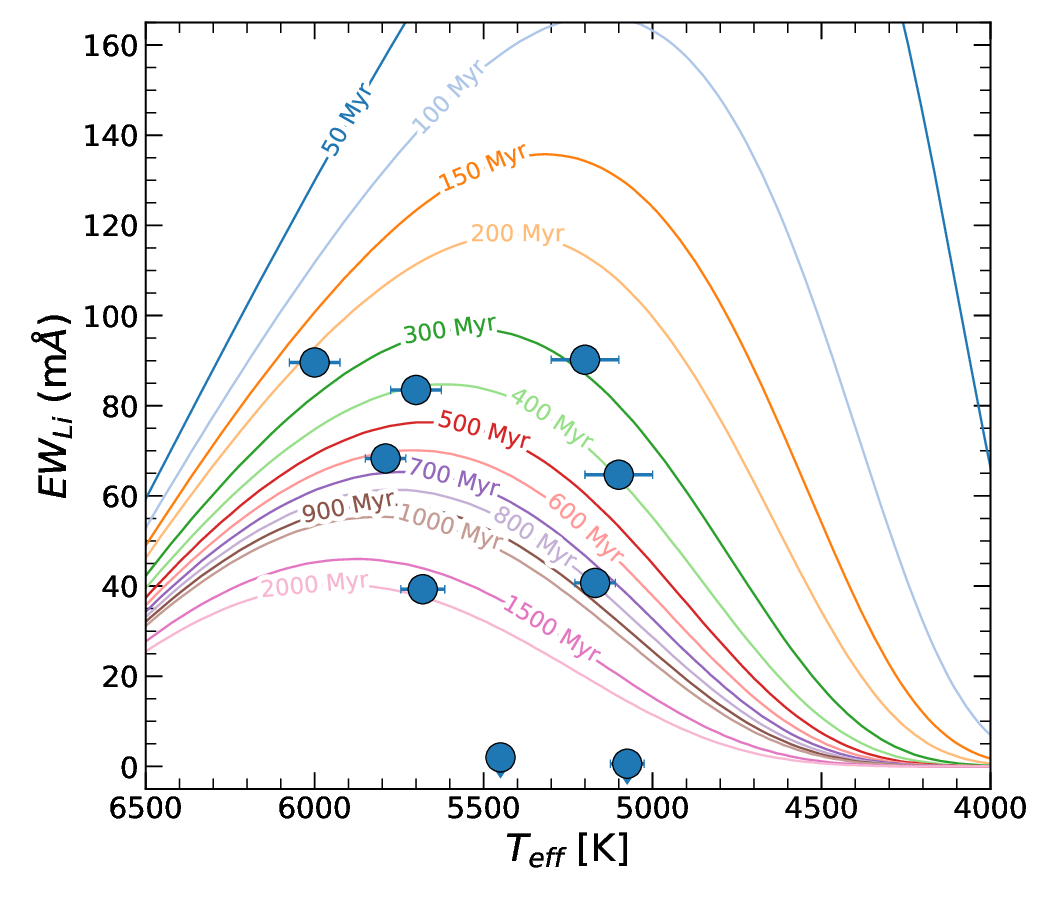}
    \caption{Lithium abundance and equivalent width vs \teff\,plots. Left: non-LTE Li abundance, $\log \epsilon {\rm (Li)}$, versus effective temperature, \teff. The targets of this work (blue filled dots) are plotted together with the targets from \cite{Biazzoetal2022} in grey open squares and the members of the Pleiades ($\sim$100 Myr; \citealt{SestitoRandich2005}), Hyades ($\sim$600 Myr; \citealt{Cummingsetal2017}), and M67 ($\sim$4.5 Gyr; \citealt{pasquini2008_m67}) clusters (in yellow crosses, red diamonds, and green triangles, respectively).
    Right: lithium equivalent width ($EW_{\rm Li}$) versus effective temperature. Here, the empirical model isochrones by \cite{jeffries_2023_eagles} at 50, 100, 150, 200, 300, 400, 500, 600, 700, 800, 900 Myr, 1, 1.5, and 2 Gyr are overplotted.}
    \label{fig:ages_eagles}
\end{figure*}

Figure \,\ref{fig:ages_eagles} shows the lithium abundance (left panel) as a function of the effective temperature, where our targets are plotted as blue dots. In the same figure, we also plot the position of members of the Pleiades ($\sim$100\,Myr, \cite{SestitoRandich2005}; yellow crosses), the Hyades ($\sim$600\,Myr, \cite{Cummingsetal2017}; red diamonds), and the M67 ($\sim$4.5\,Gyr, \cite{pasquini2008_m67}; green triangles) clusters, together with the older transiting planet-hosting stars (grey squares) with detection of the lithium line and investigated by \cite{Biazzoetal2022}. From this figure, two targets (namely, TOI-2048 and TOI-2076) are placed in the lower envelope of the Li distribution of the Pleiades, TOI-179 appears to be in between the lower envelope of the Pleiades and the Hyades cluster, other four targets (TOI-1136, TOI-1726, TOI-5082 and TOI-5398) are very close to the Hyades position, TOI-1430 seems to follow the Hyades sequence for low abundance values, while TOI-4515 is located between the Hyades and the older M67. 

We estimated the age of our targets from the lithium content considering a set of empirical isochrones in the $EW_{\rm Li}$ versus \teff\,plane calibrated using all the data for open clusters with age of 1-6000\,Myr observed as part of the \textit{Gaia}-ESO Survey (GES; \citealt{Randichetal2022}, \citealt{Gilmoreetal2022}) and implemented in the {\sc eagles} (Empirical AGes from lithium Equivalent widthS; \citealt{jeffries_2023_eagles}) code. This code models a set of stars by fitting the empirical isochrones and an intrinsic dispersion around those isochrones, for data in the range of 3000-6500\,K and $-$0.3 dex < [Fe/H] < 0.2 dex. A combined likelihood for the stars is multiplied by a prior probability to form a posterior probability distribution of age. Since it does not use lithium abundance, it is independent of the temperature scale or the adoption of any set of stellar atmosphere models or NLTE corrections. 

Table\,\ref{tab:ages_eagles} reports in the second column the most probable age and an asymmetric 68$\%$ confidence interval obtained through the {\sc eagles} tool, while the right panel of Fig.\,\ref{fig:ages_eagles} shows the lithium equivalent width as a function of the effective temperature. In this figure, our data (blue dots) are plotted over a set of model isochrones in the 50\,Myr-2\,Gyr range, as derived by \cite{jeffries_2023_eagles} through the fitting of the GES training data. 

All the targets in our sample have the $\log R'_{\rm HK}$ index measured as part of this study as well as their rotation period known from the literature or estimated in this work from TESS data with the same approach used in \cite{Messinaetal2022} (see Table \ref{tab:target_references}). 
As an independent check of the age determination, we also estimated the ages of our targets using the calibrated age-rotation and age-activity relations by \cite{mamajek_age_est_2008}.
The derived gyrochronologic and chromospheric ages are listed in Cols. 3 and 4 in Table \ref{tab:ages_eagles}. Other than TOI-5543 (Li line not deteced), TOI-1430 and TOI-4515 (for which we could measure lower limit in age through the Li line), only two stars, namely TOI-5082 and TOI-179, show ages derived from Li slightly outside 1$\sigma$ with respect to both gyrochronological and activity ages. For the remaining five stars, the three age estimation methods give similar results. We performed a two-tailed Student's test for dependent samples and found that the null hypothesis, that is, that the Li ages come from the same population of gyrochronological and activity ages, cannot be rejected (p > 0.7).

We remark here that \cite{jeffries_2023_eagles} claim that for $EW_{\rm Li}$ lower than 50\,m\AA\,the {\sc eagles} code can yield \textbf{lower limit} to ages. This is because the probability distribution becomes quite asymmetric as data points approach regions where $EW_{\rm Li}$ becomes less sensitive to age because lithium starts to be depleted. This leads to a likelihood function poorly constrained for older ($\gtsim$1 Gyr) ages. Within our sample, we have four targets with $EW_{\rm Li}<50$\,m\AA. This is the case of TOI-1430 and TOI-4515 for which we could not constrain the age because of the very low $EW_{\rm Li}$ value (0.6\,m\AA) and the $EW_{\rm Li}$ undefined mean value ($<$ 2\,m\AA), respectively. 
However, the lower limits in ages derived for these two targets using the lithium line are consistent with the recent estimates by Nardiello et al. in prep. (see also ages by \citealt{Orell-Miquel2023} and \citealt{carleoetal2024} of 700$\pm$150 Myr and 1.2$\pm$0.2 Gyr, respectively). Again, the position of these two targets in the $\log \epsilon {\rm (Li)}$-\teff\,diagram is consistent respectively with ages close to and older than the Hyades. Two other targets with $EW_{\rm Li}$ lower than 50\,m\AA\,are TOI-5082, for which no age was estimated in the literature, and TOI-179, for which the peak of the posterior probability distribution is at $\sim$900\,Myr that is older than the recent determination performed by \cite{Desidera_2023_toi179} and based on gyrochronology (400$\pm$100\,Myr). In the latter case, we also remark that the position of the target in the $\log \epsilon {\rm (Li)}$-\teff\,diagram is indeed more consistent with the age values older than the Pleiades and closer to the Hyades clusters, in agreement with the gyrochronological and activity age estimates (Table\,\ref{tab:ages_eagles}). For the other targets with $EW_{\rm Li}>50$\,m\AA, we find good agreement within 1$\sigma$ with the literature values. TOI-1726 is indeed member of the Ursa Major group ($\sim$400\,Myr; \citealt{Mannetal2020}), TOI-2048 is member of the Group X ($\sim$300\,Myr; \citealt{Messinaetal2022}), TOI-2076 is a comoving system of TOI-1807 (300$\pm$80\,Myr; \citealt{nardiello_2022}, \citealt{hedges_2021_toi2076_yo43}), and, for TOI-1136, \cite{Dai_2023_so04} found an age of $700\pm100$ Myr from gyrochonology and activity indicators. In the case of TOI-5398, we find an age from lithium abundance which is younger than that one found by \cite{Mantovanetal2023} through girochronology ($650\pm150$\,Myr), but closer to their value of $\sim 370$\,Myr derived through activity indicators. However, the higher value provided by the {\sc eagles} code (see Table\,\ref{tab:ages_eagles}) and the position in the $\log \epsilon {\rm (Li)}$-\teff\,diagram near the Hyades cluster (Fig.\,\ref{fig:ages_eagles}, left panel) seems to be more consistent with the gyrochronological age estimated in \cite{Mantovanetal2023}.

\begin{table}[h]
\renewcommand{\arraystretch}{1.5}
\caption{Age estimates obtained from different methods.}
\label{tab:ages_eagles}
\begin{center}
\begin{tabular}{lrrrr}
\hline
Name & Age$_{\rm Li}$ & Age$_{\rm Gyro}$ & Age$_{\rm Chro}$ & Age$_{\rm Lite}$ \\
 & (Myr) & (Myr) & (Myr) & (Myr) \\ \hline
TOI-179 & 895$_{-460}^{+1540}$ & 390$\pm$15 & 200$\pm$100 & 400$\pm$100$^a$ \\
TOI-1136 & 710$_{-475}^{+865}$ & 660$\pm$60 & 320$\pm$50 & 700$\pm$100$^b$ \\
TOI-1430 & >720 & 620$\pm$25 & 530$\pm$130 & 600-800$^c$ \\
TOI-1726 & 450$_{-290}^{+460}$ & 340$\pm$20 & 480$\pm$170 & 414$\pm$23$^d$ \\
TOI-2048 & 425$_{-205}^{+405}$ & 280$\pm$10 & 205$\pm$30 & 300$\pm$50$^e$ \\
TOI-2076 & 305$_{-135}^{+210}$ & 300$\pm$10 & 300$\pm$50 & 300$\pm$80$^f$ \\
TOI-4515 & >1260 & 1000$\pm$60 & 500$\pm$100 & 1200$\pm$200$^g$ \\
TOI-5082 & 2175$_{-1330}^{+3290}$ & 625$\pm$50 & 530$\pm$140 & $\dots$ \\
TOI-5398 & 285$_{-255}^{+390}$ & 520$\pm$40 & 350$\pm$100 & 650$\pm$150$^h$ \\
TOI-5543 & $\dots^*$ & 450$\pm$20 & 320$\pm$50 & $\dots$ \\ \hline
\end{tabular}
\end{center}
\begin{tablenotes}
    \small {
    \item{$^*$ No lithium line.}
    \item{Note: Col. 2 lists the most probable age from lithium equivalent width with the lower and upper bounds of the asymmetric 68 per cent confidence interval, as derived from the {\sc eagles} code \citep{jeffries_2023_eagles}. Columns 3 and 4 show the age estimates obtained from gyrochronology and chromospheric activity indicators, while Col. 5 shows age values from the literature.}
        \item{References:
 $^a$\cite{Desidera_2023_toi179}, $^b$\cite{Dai_2023_so04}, $^c$ \cite{Orell-Miquel2023}, $^d$\cite{Damasso_2023_yo40}, $^e$\cite{Newton_2023_yo42},
 $^f$\cite{nardiello_2022},
 $^g$\cite{carleoetal2024},
 $^h$\cite{Mantovanetal2023}.}
    }
\end{tablenotes}
\end{table}

\subsection{Mineralogic diagnostics}\label{sec:mineral_diagnostics}
\label{sec:mgsi_co_fe}

Abundance ratios in planet-hosting stars are useful tools to obtain important information about the composition of their planets since they govern the distribution and formation of chemical species in the protoplanetary disks where they formed. Here in the following, we discuss how some geochemical ratios derived for planet-hosting stars may become useful indicators of the planetary mantle and core compositions, in particular for low-mass planets. 

\subsubsection{The importance of the [Mg/Si] ratio}

\begin{figure}[h]
    \centering
    \includegraphics[width =\linewidth]{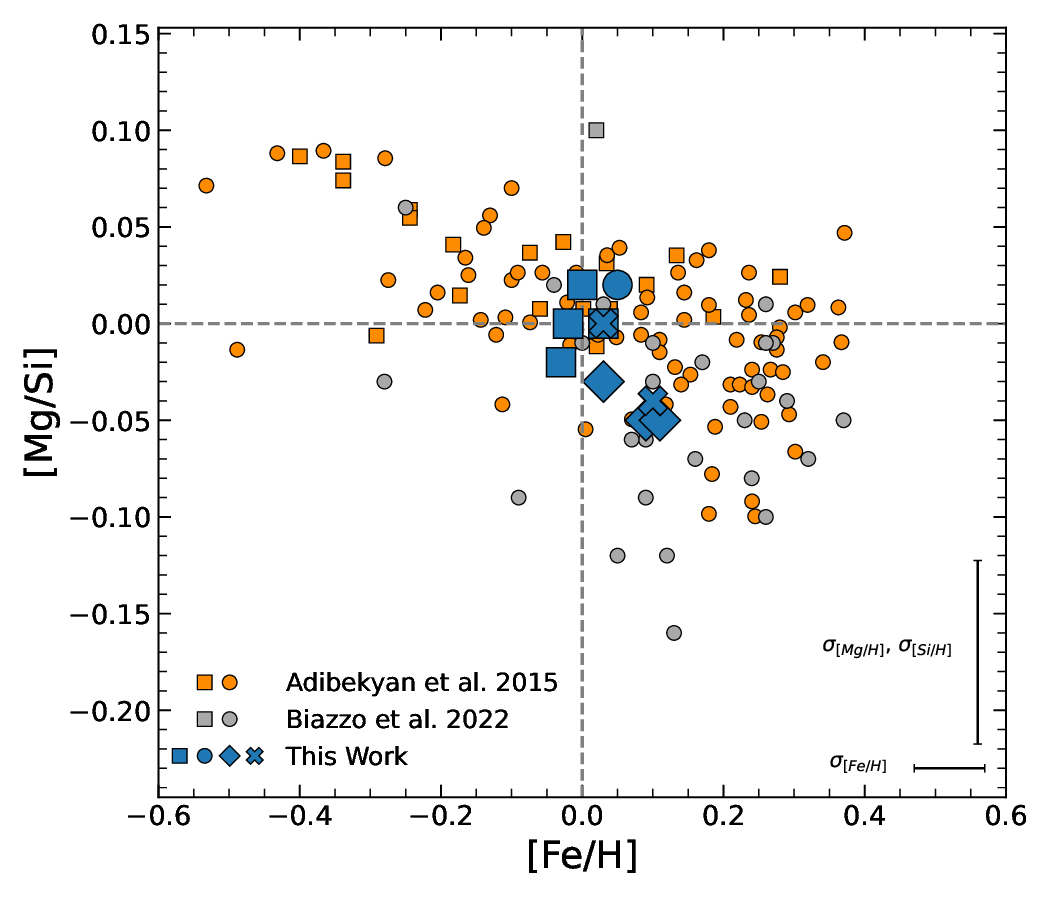}
    \caption{[Mg/Si] versus [Fe/H] for our sample (blue points), where mean errors on [Mg/H], [Si/H], and [Fe/H] are similar to each other ($\sim$0.09-0.10 dex; as a reference, typical errors in Mg, Si, and Fe abundances are represented in the bottom right corner of the plot). Overplotted in grey and orange are the results by \cite{Biazzoetal2022} and \cite{adibek2015}, respectively.
    As in \cite{adibek2015}, squares and circles represent targets hosting only planets with $M_{\rm p}$<30$M\rm_E$ (low-mass planets) and $M_{\rm p}$>30$M\rm_E$ (high-mass planets), respectively. Within our sample, we mark with diamonds those targets hosting both high- and low-mass planets and with crosses targets whose planets have not a mass estimate yet. Dashed lines represent the solar abundances.}
    \label{fig:mgsi_feh_adibek}
\end{figure}

In Fig.\,\ref{fig:mgsi_feh_adibek} we show the [Mg/Si] ratio as a function of [Fe/H]. As demonstrated by \cite{Bond_2010}, \cite{thiabaud_2015} and \cite{adibek2015}, [Mg/Si] may play an important role in the internal structure and composition of planets. In particular, \cite{adibek2015} found that this ratio can be used to constrain the chemical composition of low-mass planets and also that planets around stars belonging to the Galactic thick disk have higher values of this ratio when compared to their thin disk counterparts. This is because Mg, Si, and Fe elements have differences in the production sites within the Galactic disk and this is reflected in the observed trends. Indeed, \cite{adibek2015} found that low-mass planets are more prevalent around stars with high [Mg/Si], thus implying that the host stellar atmospheric abundances of refractory elements (such as Mg, Si, Fe) not only can they be considered a useful diagnostics of the composition of the initial protoplanetary disk, but also a proxy of the composition of very low-mass planets (\citealt{Adibekyanetal2021}). Although the relatively small sample analysed in this work is composed of only thin disk stars, we can conclude that our targets show values comparable with those by \cite{adibek2015}, with all low-mass planet-hosting stars (blue-filled squares) having [Mg/Si] around 0\,dex and higher than those of the two targets hosting both high- and low-mass planets. TOI-4515, that hosts the only gas giant planet in our sample, shows a [Mg/Si] value around 0.02\,dex, within the range between $-0.05$ and $+0.05$\,dex found for higher-mass planet-hosting stars in \cite{adibek2015} at similar [Fe/H].

\subsubsection{Key exoplanetary geochemical ratios}\label{sec:key_exo_geochem}

\begin{figure}[h]
    \centering
    \includegraphics[width=1\linewidth]{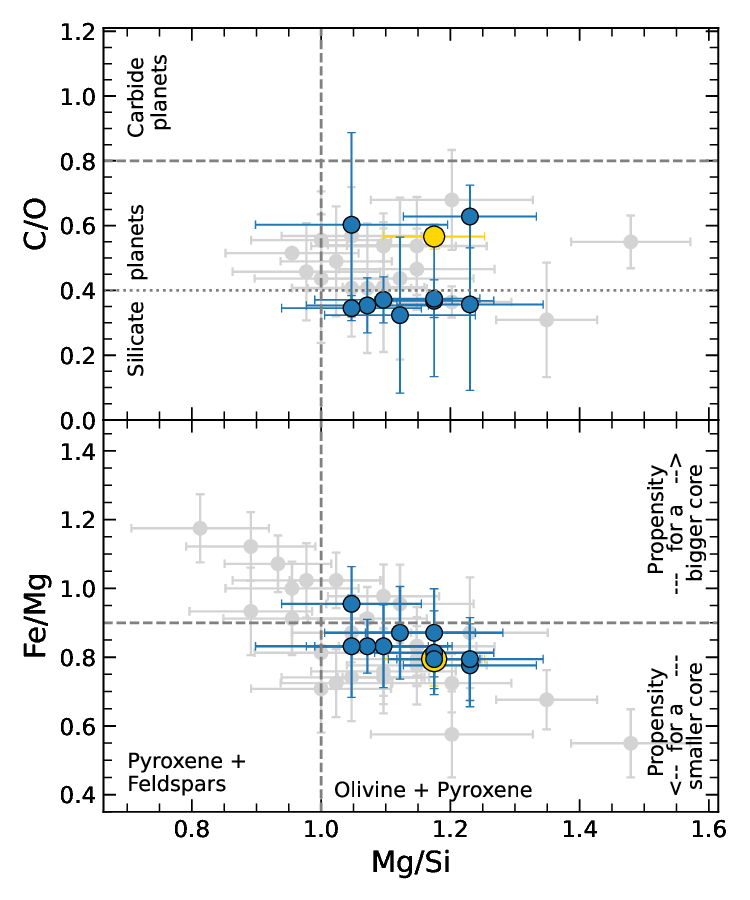}
    \caption{C/O (upper panel) and Fe/Mg (lower panel) ratios versus Mg/Si. In grey, the results by \cite{Biazzoetal2022} are shown, while the yellow big dot in both panels represents the position of the Sun as derived by the same authors (see also \citealt{Biazzoetal2023}). Vertical lines represent Mg/Si=1.0 (\citealt{thiabaud_2015}), while horizontal lines are plotted for C/O=0.4, 0.8 \citep{Suarez-andres_2018}, and for Fe/Mg=0.9 (\citealt{wang_2022}).
    }
    \label{fig:co_femg_vs_mgsi}
\end{figure}

In Fig.\,\ref{fig:co_femg_vs_mgsi} we show the distribution of our stars in the C/O\footnote{In the paper, the X$_1$/X$_2$ ratio refers to the elemental number ratio: X$_1$/X$_2$=$10^{\log \epsilon(X_1)}/10^{\log \epsilon(X_2)}$, with \logeps(X$_1$) and \logeps(X$_2$) absolute abundances.} versus Mg/Si and Fe/Mg versus Mg/Si diagrams, compared with the sample of transiting planet-hosting stars studied by \cite{Biazzoetal2022}. These elemental ratios are very useful mineralogical diagnostics because they govern the distribution and formation of chemical species in the protoplanetary disks, and, by extension, the composition of planets. On one hand, refractory elements (as traced, e.g. by the Mg/Si and Fe/Si ratios) with their high condensation temperatures ($T_{\rm cond}$) condense close to the star. Their stellar ratios, therefore, remain roughly constant throughout the planet-forming disk, thus reflecting its composition and possibly the composition of the planetary atmosphere (see \citealt{Brewer_2016}, and references therein). On the other hand, volatile elements (as traced, e.g. by C/O) have low $T_{\rm cond}$ and their ratios are governed by ices and, therefore, their values in planetary composition can deviate significantly from those of the host stars (\citealt{thiabaud_2015}). If the C/O ratio is above a threshold value (typically of $\sim$0.8) there is almost no free oxygen available to form silicates and the geology will be dominated by carbonates, while below that threshold, planetesimal geology primarily consists of magnesium silicates. In the latter case, the exact composition of silicates in the mantle of a planet depends on the Mg distribution: for Mg/Si $\ltsim$ 1, Mg forms orthopyroxene and Si is present as other silicate species such as feldspars or olivine; for 1$\ltsim$ Mg/Si $\ltsim$ 2, Mg is distributed between olivine and pyroxene; for Mg/Si $\gtsim$ 2, all available Si is consumed to form olivine with excess Mg available to bond with other minerals, mostly oxides (\citealt{Bond_2010}, \citealt{thiabaud_2015}). Further, another diagnostic, such as the Fe/Mg ratio, can be considered as an indicator of the degree of core-mantle fractionation, thus being useful for determining the core size of low-mass planets. Values of Fe/Mg greater than $\sim 1$ are interpreted as a propensity for the planets to have bigger (iron) cores (see, \citealt{wang_2022}, and references therein).

Our transiting planet host stars are all below the threshold of 0.8 in C/O and are mainly concentrated to values close to 0.4, with a mean value of <C/O>=0.43$\pm$0.12, thus consistent with silicate planets. Most of them are at the lower edge of the similar distribution done by \cite{Biazzoetal2022} for older stars. Only two targets show values slightly higher than the solar C/O ratio (yellow dot in the figure at C/O=0.57$\pm$0.04). This highlights how the remark by \cite{Turrini2022} on possible biases in the interpretation of planetary compositional models based on the use of solar values of C/O as reference stellar abundances stands also at younger evolutionary stages as those of our stars (see also \citealt{Turrini2021, Biazzoetal2022, Jorgeetal2022}). Moreover, all our targets show Mg/Si between 1.0 and 1.3, with a mean value of 1.14$\pm$0.07, only slightly lower than the solar reference of 1.17$\pm$0.08 in \cite{Biazzoetal2022} and within the errors, while the Fe/Mg ratio displayed in the lower panel of Fig.\,\ref{fig:co_femg_vs_mgsi} shows a mean value of <Fe/Mg>=0.83$\pm$0.05, which is slightly higher, within the errors, than the solar value of 0.79$\pm$0.07.

Summarising, the peak of the Mg/Si-C/O distribution for our sample is consistent with mantles of Si which will take solid form as SiO$_4^{-4}$ and SiO$_2$ and Mg made of a mixture of pyroxene and olivine assemblages, analogously to what was obtained by, for example \cite{Suarez-andres_2018} for solar-type planet-hosting stars and by \cite{Biazzoetal2022} for older transiting planet-hosting stars with similar atmospheric properties. At the same time, the peak of the Mg/Si-Fe/Mg distribution is close to the threshold where planetary cores start to be bigger (\citealt{wang_2022}).

\begin{figure}[h]
    \centering    \includegraphics[width=1\linewidth]{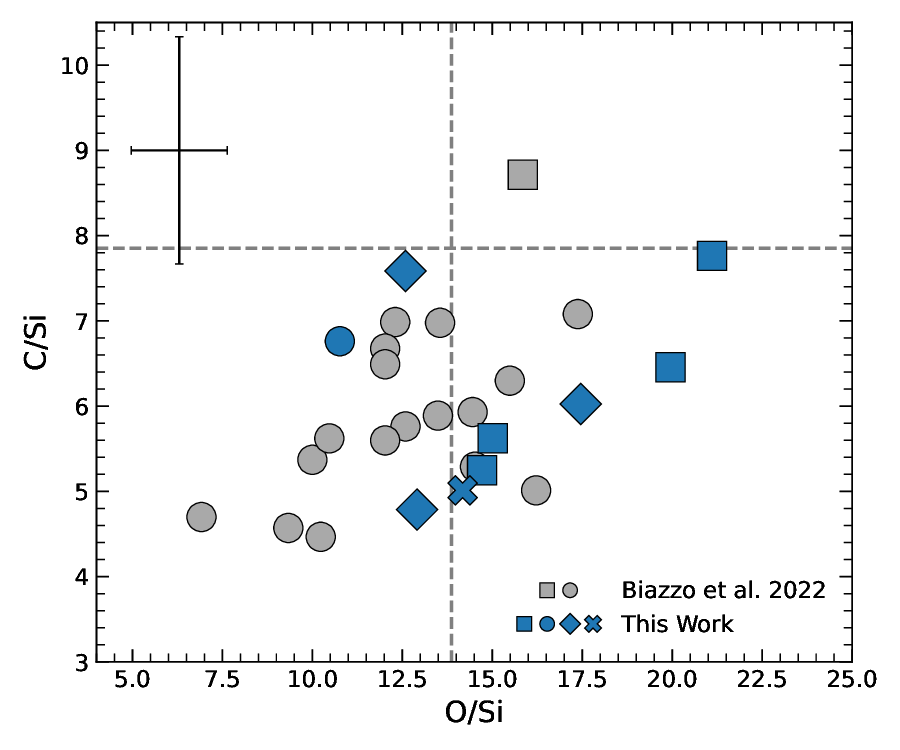}
    \caption{C/Si versus O/Si for our stars (blue points) and those (in grey) analysed by \cite{Biazzoetal2022}. Squares and circles represent targets hosting only planets with $M_{\rm p} < 30M_{\rm E}$ (low-mass planets) and $M_{\rm p} >30 M_{\rm E}$ (high-mass planets), respectively. Diamonds mark those targets hosting both higher- and lower-mass planets, while the cross symbol indicates that planet-hosting target without any planetary mass measurement. The typical uncertainty was estimated referring to the methodology used by \cite{DelgadoMena2010} and it is shown on the top left corner. Dashed lines represent the solar abundance ratios by \cite{Biazzoetal2022}.}
    \label{fig:csi_vs_osi}
\end{figure}

As already mentioned above, although the C/O ratio provides useful constraints on the potential formation locations of gaseous exoplanets, carbon and oxygen alone are not enough to determine where a given planet originated. Other ratios were therefore proposed in the literature, in particular those including refractory elements, which are present in solid phase throughout most of the disk and therefore they reflect the solid-to-gas accretion ratio of the material incorporated into the planetary envelope during the formation (see \citealt{Turrini2021}). Very recently, \cite{Chachan_2023} proposed O/Si and C/Si as possible ratios useful to alleviate the degeneracies in the interpretation of atmospheric C/O ratios, making it easier to relate the planetary abundance ratios to the compositions of the solids in the disk. In Fig.\,\ref{fig:csi_vs_osi} we show the C/Si versus O/Si ratios for our targets and those analysed by \cite{Biazzoetal2022}. All targets are below the solar value in the C/Si ratio (with the exception of one target investigated by \citealt{Biazzoetal2022}; see that paper for details), consistently with the statements by \cite{Chachan_2023} for planet-hosting stars with high content of refractory elements. At the same time, both samples of transiting planet-hosting stars show a wide distribution in O/Si, with a propensity for lower-mass planet-hosting stars to possess higher values of O/Si. Considering the planet-hosting stars homogeneously analysed in this work and in \cite{Biazzoetal2022}, the mean O/Si for both samples is indeed of $17.3\pm3.0$ and $12.7\pm2.7$ for stars hosting planets with $M_{\rm p} < 30M_{\rm E}$ and $M_{\rm p} > 30M_{\rm E}$, respectively. By chance, all the five targets hosting only planets with $M_{\rm p} < 30M_{\rm E}$ appear to have super-solar O/Si ratio. This seems to be again consistent with the claim by \cite{Chachan_2023} for which hot Jupiter host stars should show lower content of O/Si and host planets with silicate clouds and low water content. The possible relation between stellar O abundance and planetary mass will be discussed in Sect.\ref{sec:abundances_plmass}.

Similar results were found by analysing the C/S ratio versus O/S, which were recently proposed by \cite{Crossfield_2023} as useful diagnostics of the formation of warm and hot giant planets. In the C/S vs O/S relation, shown in Fig.\ref{fig:cs_vs_os}, we used the same symbols as in Fig.\ref{fig:csi_vs_osi} to distinguish targets hosting low- and high-mass planets. Similarly to the C/Si vs O/Si distribution, using sulphur S as refractory element we observe a tendency for lower-mass planet-hosting stars to be present at higher values of O/S. The mean O/S values for stars hosting planets with $M_{\rm p} < 30M_{\rm E}$ and $M_{\rm p} > 30M_{\rm E}$, are indeed 37.4$\pm$3.2 and 30.7$\pm$9.0 respectively.

We are aware that these are presumably speculations because at least part of the trends could be due to selection effects (e.g. stellar age and atmospheric parameters), but we think that these types of possible relationships that include multiple refractory and volatile elemental ratios should be studied widely and systematically in the next-coming and future big spectroscopic surveys.

\subsubsection{Bulk composition estimates of low-mass planets}

Due to the above-mentioned limitations in the use of stellar volatile abundances to make assumptions on planetary composition, \cite{wang_2019} introduced the idea of using devolatilised host stellar abundances to constrain the bulk composition and interior modelling of hypothetical habitable-zone terrestrial exoplanets. This approach is limited to this kind of planets since it is based on the observations of bulk composition differences and similarities of the Solar System's rocky bodies relative to the Sun. Among the major rock-forming elements analysed by the authors, only Ca and Al are not observed to be depleted in rocky bodies relative to the Sun, while for volatiles like O, S, and C the depletion is over 80\%. Therefore, in rocky planets, elements like C and O are no longer valid indicators of the mantle oxidation state, which however is essential to understand the planetary interiors. As a consequence, the authors proposed the bulk (O$-$Mg$-$2Si)/Fe as a diagnostics of the oxygen fugacity, which is an indicator of the mantle oxidation state of a silicate terrestrial exoplanet, with a depletion of Mg, Fe, and Si of around 15\% (see \citealt{wang_2019} for details).

\begin{figure*}[t!]
    \centering
    \includegraphics[width=1.\textwidth]{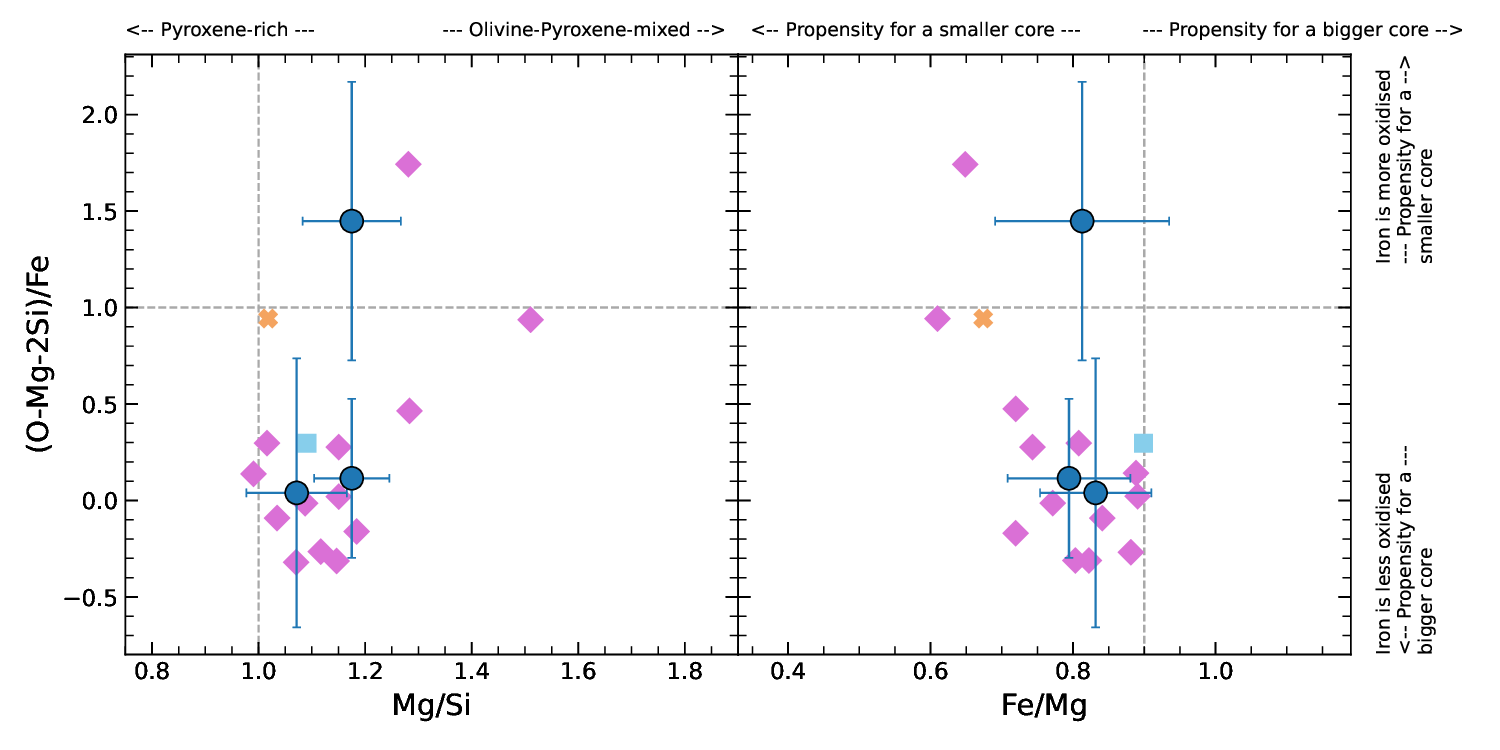}
    \caption{(O$-$Mg$-$2Si)/Fe versus Mg/Si (left panel) and (O$-$Mg$-$2Si)/Fe versus Fe/Mg (right panel) diagrams. The vertical and horizontal dashed lines (\citealt{wang_2022}) highlight the areas in which planets are expected to have different planetary interior properties, along with the positions of Mars (cross) and Earth (square) shown for reference (see \citealt{wang_2022}). Overplotted with diamonds are the terrestrial-type exoplanets analysed by \cite{wang_2022}.
    }
    \label{fig:wang_OMgSi_Fe}
\end{figure*}

Figure\,\ref{fig:wang_OMgSi_Fe} shows the distribution of our targets in the (O$-$Mg$-$2Si)/Fe versus Mg/Si and (O$-$Mg$-$2Si)/Fe versus Fe/Mg diagrams, where we applied the devolatilisation model to only those stars hosting planets with mass lower than 10\,$M_{\rm E}$ (or with radius lower than 3\,$R_{\rm E}$ in case of TOI-2048 and TOI-5082 for which no planetary mass estimate was obtained by the time of this publication). In our sample, only four targets host planets satisfying the mass (or radius) requirements (namely TOI-1136, TOI-1430, TOI-2048, and TOI-5082). Figure\,\ref{fig:wang_OMgSi_Fe} considers the Mg/Si ratio as an indicator of the dominant mineral assemblages (i.e. olivine or pyroxene) in the mantle of a silicate planet. In the same figure, the Fe/Mg ratio is used as a proxy of the core-mantle fractionation degree, hence the core size (see details in \citealt{wang_2022}).

The dashed lines divide our low-mass planets into different categories, in which different planetary interior properties may be expected, as indicated by \cite{wang_2022}. Two of our four sample planets (orbiting TOI-1136 and TOI-5082) are below the unit line of (O$-$Mg$-$2Si)/Fe, implying a potential large iron core. These two planets with (O-Mg-2Si)/Fe lower than 1 show Fe/Mg values consistent within the errors with the Earth value (square symbol in Fig.\,\ref{fig:wang_OMgSi_Fe}), thus their mantle mineralogies would be more or less the same, as also confirmed by their similar values in the Mg/Si ratio.
The iron in the low-mass planet around TOI-1430, with (O$-$Mg$-$2Si)/Fe > 1, may be more oxidised and locked in the mantle, thus developing comparably smaller core. 
TOI-2048 is missing in the plot because it has no oxygen abundance estimate (as shown in Table\,\ref{tab:oth_elem_abund}).

\subsection{Stellar abundances versus planetary properties}
\label{sec:abundances_plmass}

\begin{figure}[h!]
    \centering
    \includegraphics[width=1\linewidth]{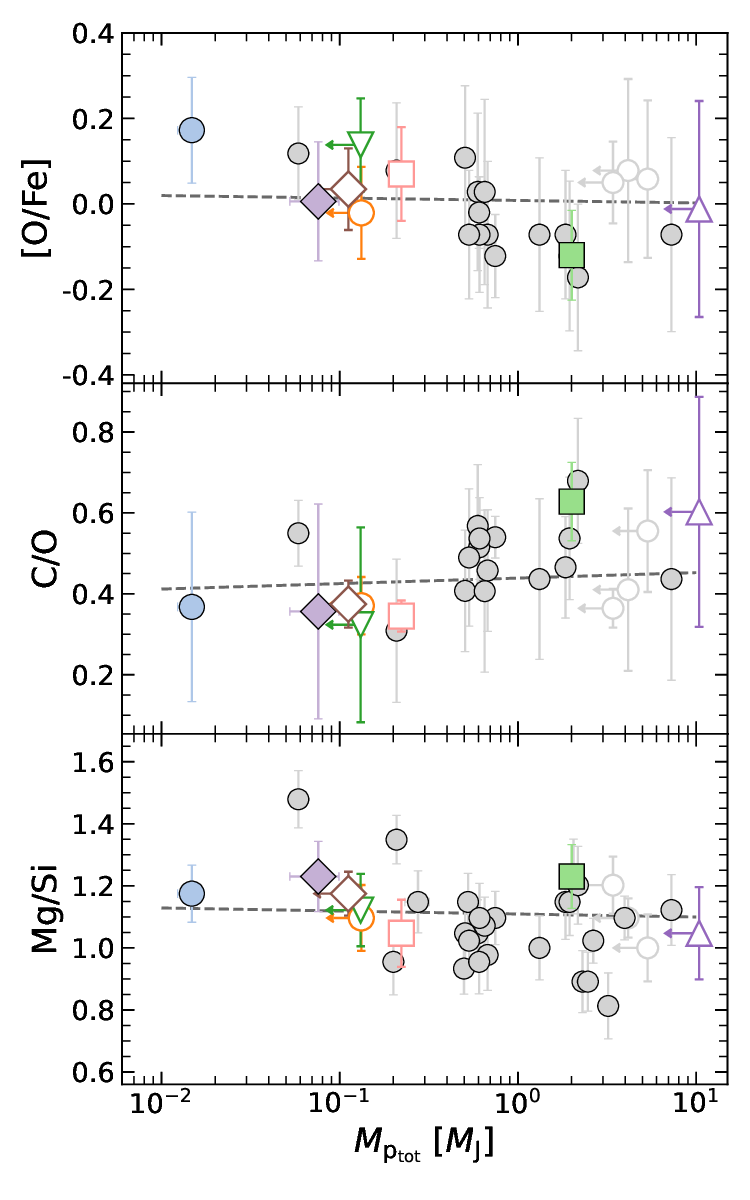}
    \caption{[O/Fe] as well as C/O and Mg/Si elemental abundance ratios as a function of the total planetary mass of the planetary system. Filled and empty symbols represent the targets that within our sample host only one confirmed transiting planet or multi-planetary systems, respectively. The targets by \cite{Biazzoetal2022} are overplotted in grey circles.
    The dashed grey lines represent the linear fits on the data points.
    }
    \label{fig:_o_fe_co_mgsi_vs_mp}
\end{figure}

With the aim to look for possible relations between stellar abundances and planetary parameters, we analysed the [X/Fe] ratios for the volatile elements measured in the present work (namely, C, O, and S) as a function of the the total planetary mass of the planetary system ($M_{\rm p_{tot}} = \sum_i^{n_{planets}} M_{\rm p_i}$), which we adopt as a proxy into the planet formation efficiency of the original protoplanetary disks of the stars in the sample.
As in \cite{suarezandresetal2017} and \cite{Biazzoetal2022}, we do not find a significant trend with C and S, while a notable correlation seems to be present for the [O/Fe] ratio versus $M_{\rm p_{tot}}$ (shown in Fig.\,\ref{fig:_o_fe_co_mgsi_vs_mp}, top), with a decreasing trend towards greater masses. In particular, defining the significance \textit{p}-value threshold $\alpha=0.05$ (where a statistically significant relation has \textit{p}<$\alpha$), we find a Spearman statistical significance of $p \sim 0.01$ for the oxygen decreasing trend with $M_{\rm p_{tot}}$. For carbon and sulphur, the relation is not significant ($p \sim 0.65-0.93$), with the [S/Fe] distribution showing higher dispersion.

The observed trend between [O/Fe] and the total planetary mass has an intuitive physical interpretation when we look into the role of O in shaping the mass of heavy elements in stars and the amount of planet-forming material in protoplanetary disks. Specifically, O is the heavy element that contributes most mass to the stellar metallicity (about 45\% in the case of the solar composition, \citealt{Lodders2010}) and comparative studies between the solar composition and meteoritic abundances reveal that, mass-wise, it is a key component in the planetary building blocks already from the inner regions ($\sim$ few au) of protoplanetary disks. In the Solar Nebula up to half the protosolar budget of O was incorporated into rocks and dust before the water snowline \citep{Lodders2010} while a large fraction of the remaining O condensed as water ice immediately after \citep{Oberg2021,Turrini2021}. Studies of the interstellar medium, circumstellar disks and polluted white dwarfs suggest that this key role of O is general to the Solar neighbourhood \citep{Oberg2021,Trierweiler2023}. Increases in the abundance of O should therefore translate into proportional increases in the mass of solid material that can support the growth of planetary bodies and enhance the planet formation efficiency of protoplanetary disks.

To test this we performed the following population synthesis study with our Monte Carlo implementation of the GroMiT code \citep{Polychroni2023} previously used in \citet{Mantovan2024}. The code implements the prescriptions for the growth and migration of solid planets by pebble accretion from \citet{Johansen2019} and those for the growth and migration of gas-accreting giant planets by \citet{Tanaka2020}, complementing them with the treatment of the solids-to-gas ratio in the protoplanetary disk from \cite{Turrini2023}. We simulated the formation of synthetic planetary populations in two identical protoplanetary disks differing only for their O abundances and compared the resulting planets. The first disk was characterised by solar composition and metallicity (1.4\%; \citealt{Asplund2009}), while the second disk was enriched in O by 0.3 dex, resulting in a supersolar metallicity of 2.1\%. Since the amount of O sequestered by rocks is controlled by the abundances of Fe, Mg and Si \citep{Burrows1999,Fegley2010} and these elements have the same abundances in both disks, the larger abundance of O in the O-enriched disk should affect only the amount of solids beyond the water snowline.

To correctly reproduce this effect, we modified the prescription for the solid-to-gas ratio from \citep{Turrini2023} between the two disks. Specifically, we assume both disks being characterised by the same solid-to-gas ratio in their inner and hotter regions dominated by rocks, with the O-enriched disk having a larger solid-to-gas ratio beyond the water snowline to represent its larger budget of water ice. As a result, in the solar disk the solid-to-gas ratio inside the water snowline is set to 0.7\% (0.5x the disk metallicity, \citealt{Turrini2023}) and is 1.05\% (0.75x the disk metallicity, \citealt{Turrini2023}) beyond the water snowline based on the case of the Solar System \citep[see][for a detailed discussion]{Turrini2021}. In the O-enriched disk, the solid-to-gas ratio inside the water snowline is kept to 0.7\% and is increased to 1.58\% beyond the water snowline.

\begin{table}
    \caption{Initial parameters used to run the modified GroMiT code. }
    \label{tab:popsythesis}
    \centering
    \begin{tabular}{l c c}
        \hline \hline
    \multicolumn{2}{c}{Simulation Parameters} \rule{0pt}{2.3ex} \rule[-1ex]{0pt}{0pt}\\
    \hline
      \multicolumn{1}{l}{N$^\circ$ of seeds per disk} & \multicolumn{1}{c}{5$\times$10$^4$} \rule{0pt}{2.3ex} \rule[-1ex]{0pt}{0pt}\\
      \multicolumn{1}{l}{Seed formation time} & \multicolumn{1}{c}{0.01--2$\, \times \, 10^6$\,yr} \\
      \multicolumn{1}{l}{Disk lifetime} & \multicolumn{1}{c}{$5 \times 10^6$\,yr} \\
    \hline
    \multicolumn{2}{c}{Star, Planet \& Disk properties} \rule{0pt}{2.3ex} \rule[-1ex]{0pt}{0pt}\\
    \hline
      \multicolumn{1}{l}{Stellar Mass} & \multicolumn{1}{c}{1.109$\,$M${_\odot}$} \rule{0pt}{2.3ex} \rule[-1ex]{0pt}{0pt}\\
      \multicolumn{1}{l}{Seed Mass} & \multicolumn{1}{c}{0.01 $M_{\rm E}$} \\
      \multicolumn{1}{l}{Initial envelope mass} & \multicolumn{1}{c}{0.0 $M_{\rm E}$} \\
      \multicolumn{1}{l}{Initial semimajor axis} & \multicolumn{1}{c}{0.5--50$\,$au} \\
      \multicolumn{1}{l}{Disk characteristic radius} & \multicolumn{1}{c}{50$\,$au} \\
      \multicolumn{1}{l}{Temperature @ 1$\,$au} & \multicolumn{1}{c}{200$\,$K} \\
      \multicolumn{1}{l}{Surface density @ 1$\,$au} & \multicolumn{1}{c}{420\,kg m$^{-2}$} \\
      \multicolumn{1}{l}{Disk accretion coefficient, $\alpha$} & \multicolumn{1}{c}{0.01}\\
      \multicolumn{1}{l}{Dust-to-gas ratio}& {0.021 \& 0.014}\\
      \multicolumn{1}{l}{Pebble size}& \multicolumn{1}{c}{1$\,$mm}\\
    \hline
  \end{tabular}
\end{table}

For each disk we simulated the growth and migration of $5\times10^4$ planetary seeds uniformly distributed in semimajor axis across the disk extension. The initial and boundary conditions of the two population synthesis simulations are summarised in Table \ref{tab:popsythesis}. From the resulting planetary populations we extracted all planets with mass greater than about 5 $M_{\rm E}$ and divided them over two classes based on those considered in \cite{Biazzoetal2022} and in this work: massive planets between 30 and 954 $M_{\rm E}$ (roughly 0.1-3 $M_{\rm J}$) and smaller planets with masses between 5 and 30 $M_{\rm E}$. The higher metallicity of the O-enriched disk (2.1\% vs 1.4\%) makes it 1.7$\times$ more efficient in forming planets above 5 $M_{\rm E}$ than the solar disk, in line with previous studies on the role of disk metallicity on planet formation efficiency \citep{Savvidou2023}.

\begin{table}
\caption{Results from the implemented GroMiT code for the O-enriched disk and the solar disk, and divided for the planet mass ranges (5-30 $M_{\rm E}$ and 30-954 $M_{\rm E}$) and for their formation period ($<$ 1 Myr and $>$ 1 Myr).}
\label{tab:planets}
\centering
\renewcommand{\arraystretch}{1.5}
\begin{tabular}{cccc}
\hline
 & Type of Planet & O-Enriched Disk & Solar Disk \\ \hline
\multirow{2}{*}{\makecell{N$^\circ$ of \\ planets \\ formed}} & 5--30\,$M_{\rm E}$ & 5931 & 2413 \\
 & 30--954\,$M_{\rm E}$ & 10974 & 7604 \\ \hline
\multirow{2}{*}{\makecell{Relative \\ Abundance}} & 5--30\,$M_{\rm E}$ & 35\% & 24\% \\
 & 30--954\,$M_{\rm E}$ & 65\% & 76\% \\ \hline
\multirow{2}{*}{{\makecell{5--30 \\ $M_{\rm E}$}}} & <\,1\,Myr & 70\% & 76\% \\
 & >\,1\,Myr & 30\% & 24\% \\ \hline
\multirow{2}{*}{{\makecell{30--954 \\ $M_{\rm E}$}}} & <\,1\,Myr & 56\% & 60\% \\
 & >\,1\,Myr & 44\% & 40\% \\ \hline
\end{tabular}
\end{table}

In the O-enriched disk the formation of the smaller planets is boosted more (2.5$\times$ with respect to the solar disk) than that of the more massive planets (1.4$\times$ with respect to the solar disk). In the O-enriched disk the smaller planets represent 35\% of the extracted population, while in the solar disk they represent only 24\% (see Table \ref{tab:planets}). We further divided each class of planets into two sub-classes: early-born planets formed in the first 1\,Myr and late-born planets formed beyond 1\,Myr. In the O-enriched disk there is a global increase in the relative abundance of late-born planets, with the abundance of late-born planets between 5 and 30 $M_{\rm E}$ increasing twice as much as that of late-born planets above 30 $M_{\rm E}$. These results show that O-rich stars have a higher probability of forming planets between 5 and 30 $M_{\rm E}$ and support the interpretation provided here and in \citet{Biazzoetal2022} to the observed trends between [O/Fe] and the total planetary mass. In particular, the large contribution of O to the mass fraction of heavy elements in stars and their disks appears effective in promoting the formation of low-mass planets since the earlier stages of their circumstellar disk lifetime.

In the central and bottom plots of Fig.\,\ref{fig:_o_fe_co_mgsi_vs_mp}, we show the  C/O and Mg/Si ratios calculated in the present work as a function of the total planetary mass, respectively. As done for the abundances of volatile elements, we searched for possible trends between C/O and Mg/Si ratios and the total planetary masses.  
We find hints of a relation describing a decreasing trend with a rank correlation coefficient of $\rho \sim -0.27$
and Spearman statistical significance $p \sim 
0.12$ for Mg/Si. Instead, for the C/O ratio we obtain $p \sim 0.007$ and a rank correlation coefficient of $\rho \sim 0.52$ showing an increasing trend with $M_{\rm p_{tot}}$. Some evidence of a downward trend of Mg/Si-$M_{\rm p_{tot}}$ (higher values of Mg/Si for lower-mass planets) and a slightly increasing trend of C/O-$M_{\rm p_{tot}}$ (higher values of C/O for higher-mass planets) were also reported by \cite{Misheninaetal2021}, but no conclusion was drawn because of the large scatter of the elemental ratios as a function of planetary masses.

\section{Conclusions}
\label{sec:conclusions}

In this paper, we presented the analysis of a sample of ten GK TESS stars with \teff\,> 5000\,K, \vsini\,< 10\,km/s, and ages ranging between 0.2 and 2\,Gyr placed in the solar neighbourhood and hosting at least one confirmed transiting planet. This study was conducted in the framework of the GAPS sub-programme focussed on searching for exoplanets around young objects. We performed an accurate and homogeneous characterisation of relatively young planet-hosting stars from high-resolution spectra acquired with the HARPS-N spectrograph at the Telescopio Nazionale {\it Galileo}. We derived the kinematic properties, atmospheric parameters, and abundances of 18 elements. In the following, we illustrate our main results:

\begin{itemize}
    \item The chemical and kinematic analysis allowed us to confirm that our targets belong to the Galactic thin disk.

    \item We estimated stellar ages through the lithium line and the use of empirical isochrones, thus confirming the youth of our sample and the compatibility for most of the targets with the ages derived from gyrochronological and activity indicators, and those available in the literature.
    
    \item We find that all our targets show a C/O ratio smaller than 0.8, with most of them being below the solar value of $\sim$0.57 and in the lower edge of older stars that host transiting planets. This result, together with the mean Mg/Si ratio being around 1.14, suggests that our sample of planet-hosting stars is consistent with silicate mantles with Mg mostly distributed between olivines and pyroxenes. The Fe/Mg ratio indicates a propensity for bigger planetary cores.

    \item From C/Si and O/Si ratios, we observed a prevalence for transiting planet-hosting stars with a high O/Si ratio to host low-mass planets. Meanwhile, the C/Si ratio is sub-solar and consistent with a high content of refractory elements.

    \item We applied a devolatilisation model and used the bulk (O-Mg-2Si)/Fe as diagnostics of the oxygen fugacity for terrestrial planets. We find that two out of four very low-mass (<10 $M_{\rm E}$) planets show (O-Mg-2Si)/Fe values compatible with the possible presence of large iron cores and Fe/Mg and Mg/Si consistent with Earth's values.

    \item Concerning volatiles, we confirm the tendency previously found in old transiting planet hosts in which the oxygen element seems to be higher for stars hosting lower-mass companions. This possible trend seems to extend also to relatively young systems the recent suggestion that was reported in the literature according to which the formation of low-mass planets is favoured at higher stellar oxygen values.
    
\end{itemize}

Thanks to \textit{JWST} and the future launch of \textit{Ariel} in 2029, it will be possible to probe exoplanet atmospheres of young transiting planets in detail, allowing for insight into the atmospheric composition and the possible presence of clouds and hazes. This will also allow us to get insight into the formation and migration mechanisms of young exoplanets that are expected to be better understood with the future findings of the dedicated ELT instruments (such as ANDES).

\begin{acknowledgements}
We are very grateful to the referee Haiyang Wang for providing insightful and stimulating comments that
have helped us to improve the presentation of our results. This research has made use of the Exoplanet Follow-up Observation Programme (ExoFOP; DOI: 10.26134/ExoFOP5) website, which is operated by the California Institute of Technology, under contract with the National Aeronautics and Space Administration under the Exoplanet Exploration Programme, and also of the INSPECT database, version 1.0 (www.inspect-stars.net). This work has made use of the SIMBAD database and the VizieR catalogue access tool operated at CDS (Strasbourg, France), and also of the Hypatia Catalog Database, an online compilation of stellar abundance data, which was supported by NASA's Nexus for Exoplanet System Science (NExSS) research coordination network and the Vanderbilt Initiative in Data-Intensive Astrophysics (VIDA). This work has been financially supported by the PRIN-INAF 2019 “Planetary systems at young ages” (PLATEA) and by the grant INAF 2022 TRAME@JWST (TRacing the Accretion Metallicity rElationship with NIRSpec@JWST). DP acknowledges the support from the Istituto Nazionale di Oceanografia e Geofisica Sperimentale (OGS) and CINECA through the program ``HPC-TRES (High Performance Computing Training and Research for Earth Sciences)'' award number 2022-05 as well as the support of the  ASI-INAF agreement n 2021-5-HH.1-2022. GMa acknowledges support from CHEOPS ASI-INAF agreement n. 2019-29-HH.0.

\end{acknowledgements}

\bibliographystyle{aa} 
\bibliography{Biblio}

\begin{appendix}

\section{Additional figure}

\begin{figure}[h]
       \centering
    \includegraphics[width=\linewidth]{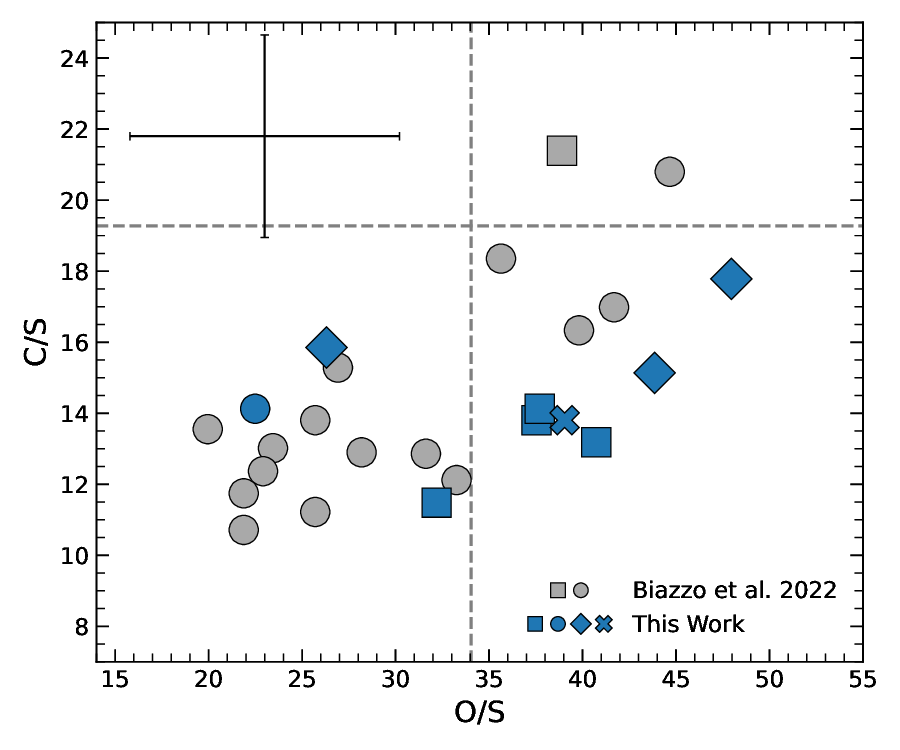}
    \caption{C/S versus O/S for our stars (blue points) and those (in grey) analysed by \cite{Biazzoetal2022}. Symbols as in Fig.\ref{fig:csi_vs_osi}. Dashed lines represent the solar abundance ratios by \cite{Biazzoetal2022}. The typical uncertainty was estimated referring to the methodology used by \cite{DelgadoMena2010} and it is represented on the top left corner of the plot.
    }
    \label{fig:cs_vs_os}
\end{figure}

Following the analysis in Sec.\ref{sec:key_exo_geochem}, Fig.\ref{fig:cs_vs_os} shows the relation between the diagnostic elemental ratios C/S and O/S, with the symbols used to distinguish targets hosting low- and high-mass planets as in Fig.\ref{fig:csi_vs_osi}.

\end{appendix}

\end{document}